\documentclass[a4paper,10pt,nofootinbib,eqsecnum,showpacs]{revtex4}
\pdfoutput=1

\usepackage{graphicx}
\usepackage{subfigure}
\usepackage{amssymb, amsmath}
\def\H{\mathcal{H}}
\newcommand{\pd}[2]{\ensuremath{\frac{\partial #1}{\partial #2}}}
\def\U{U_{0}}
\def\Uphi{U_{,\varphi}}
\def\Upp{U_{,\varphi\varphi}}
\def\Uppp{U_{,\varphi\varphi\varphi}}
\newcommand{\etal}{\emph{et al.\ }}
\newcommand{\eg}{{e.g.\ }}
\newcommand{\ie}{{i.e.\ }}
\def\Mpl{M_{\mathrm{PL}}}
\def\Mpc{M \mathrm{pc}}

\def\kmax{k_{\mathrm{max}}}
\def\kmin{k_{\mathrm{min}}}

\def\kwmap{k_\mathrm{WMAP}}
\def\fnl{f_{\mathrm{NL}}}
\newcommand{\e}[1]{{\times 10^{#1}}}
\newcommand{\lambdaphifour}{\frac{1}{4}\lambda\varphi^4}
\newcommand{\msqphisq}{\frac{1}{2} m^2 \varphi^2}

\def\vp{{\varphi}}
\newcommand\dvp[1]{{\delta\varphi_{#1}}}
\def\wt{\widetilde}
\def\kvi{{{k^i}}}
\def\qvi{{{q^i}}}
\def\pvi{{{p^i}}}
\newcommand\eq[1]{Eq.~(\ref{#1})}
\newcommand\eqs[1]{Eqs.~(\ref{#1})}
\newcommand\Rref[1]{Ref.~\cite{#1}}
\def\be{\begin{equation}}
\def\ee{\end{equation}}
\def\bea{\begin{eqnarray}}
\def\eea{\end{eqnarray}}

\usepackage[pdftitle={Numerical calculation of second order perturbations},
            pdfauthor={Ian Huston, Karim A. Malik}, 
            pdftex, hyperfootnotes=false]{hyperref}

\begin{document}
\title{Numerical calculation of second order perturbations}
\author{Ian Huston}
\email{i.huston@qmul.ac.uk}
\author{Karim A.~Malik}
\email{k.malik@qmul.ac.uk}
\affiliation{
Astronomy Unit, School of Mathematical Sciences, Queen Mary University
of London, Mile End Road, London, E1 4NS, United Kingdom
}
\date{\today}
\begin{abstract}
We numerically solve the Klein-Gordon equation at second order in
cosmological perturbation theory in closed form for a single scalar
field, describing the method employed in detail. We use the slow-roll
version of the second order source term and argue that our method is
extendable to the full equation. We consider two standard single field
models and find that the results agree with previous calculations
using analytic methods, where comparison is possible. Our procedure
allows the evolution of second order perturbations in general and the
calculation of the non-linearity parameter $\fnl$ to be examined in
cases where there is no analytical solution available.
\end{abstract}
\pacs{98.80.Cq, 98.80.Jk \hfill  arXiv:0907.2917 }

\maketitle
\section{Introduction}
Cosmological perturbation theory is an essential tool for the analysis
of cosmological models, in particular as the amount of observational
data continues to increase. With the recent launch of the
{\sc{planck}} satellite, the {\sc{wmap}} mission reaching its eighth
year, and a host of other new experiments, we will have access to more
information about the early universe than ever before
\cite{planck,Komatsu:2008hk}.

To distinguish between theoretical models 
it is necessary to go beyond the standard statistical analyses that
have been so successful in the recent past. As a result much interest
has been focused on non-gaussianity as a new tool to help classify and
test models of the early universe. Perturbation theory beyond first
order will be required to make the best possible use of 
the data.  In this paper we outline an important step in the
understanding of perturbation theory beyond first order, demonstrating
that second order perturbations are readily amenable to numerical
calculation, even on small and intermediate scales inside the horizon.

Inflationary model building has for the past few years focused on
meeting the requirements of first order perturbation theory, namely
that the power spectra of scalar and tensor perturbations should match
that observed in the Cosmic Microwave Background (CMB).  Inflationary
models are classified and tested based on their predictions for the
power spectrum of curvature perturbations, the spectral index of these
perturbations and the ratio of tensor to scalar perturbations.  As the
potential for moving beyond first order perturbations has been
explored, these three observable quantities have been joined by a
measure of the departure from gaussianity exhibited by the
perturbations, the non-gaussianity parameter $\fnl$. This parameter is
not yet well constrained by observational data in comparison with the
other quantities but can already be used to rule out models with
particularly strong non-gaussian signatures.

There are two main approaches to studying higher order effects and
non-gaussianity.  
One approach uses nonlinear theory and a gradient expansion in various
guises, either explicitly, \eg
Refs.~\cite{Salopek:1990jq,Rigopoulos:2005xx} or in the form of the
$\Delta N$ formalism, \eg
Refs.~\cite{Starobinsky:1982ee,
Starobinsky:1986fxa, Sasaki:1995aw, Sasaki:1998ug, Lyth:2004gb,Lyth:2005fi,Langlois:2006vv}
By virtue of having to employ a gradient expansion this approach is so
far only usable on scales much larger than the particle horizon.  The
other approach uses cosmological perturbation theory following Bardeen
\cite{Bardeen:1980kt} and extending it to second order,
e.g.~Refs.~\cite{Tomita:1967,Mukhanov:1996ak,Bruni:1996im,
  Acquaviva:2002ud,Nakamura:2003wk,Noh:2004bc,
  Bernardeau:2002jy,Maldacena:2002vr,
  Finelli:2003bp,Bartolo:2004if,Enqvist:2004bk,Lyth:2005du,Seery:2005gb,
  Malik:2003mv, Barnaby:2006cq}
(for an extensive list of references and a recent review on these
issues see Ref.~\cite{Malik:2008im}).
This approach works on all scales, but can be more complex than in
particular the $\Delta N$ formalism. Both these approaches give the
same results on large scales \cite{Malik:2005cy}. We
will follow the Bardeen approach in this paper.

As the first order perturbations of the inflaton field are taken in
the standard treatment to be purely gaussian it is in general
necessary to go to second order in order to understand and estimate
the non-gaussian contribution of any inflationary model (for a recent
review see Ref.~\cite{Malik:2008im}). Deriving the equations of motion is
not trivial at second order and only recently was the Klein-Gordon
equation for scalar fields derived in closed form, taking into account
metric backreaction \cite{Malik:2006ir}. This allows for the first time
a direct computation of the second order perturbation in full, in
contrast with previous attempts which have focused only on certain
terms in the expression, for example \Rref{Finelli:2006wk}.

In this paper we 
solve numerically the second order Klein-Gordon
equation in closed form in Fourier space and show that this procedure
is readily applicable to the study of non-gaussianity and other higher
order effects.
As this is, to our knowledge, the first numerical solution to the full
second order evolution equation we will outline the numerical steps
taken in the system we have developed, examine the current constraints
on the calculation and describe the next steps required in
detail. This calculation uses the slow roll version of the second
order equation, but solves the full non-slow roll equations for the
background and first order.
The models that we test in this paper are single field models with a canonical
action. Significant second order corrections
are expected only when a non-canonical action or multiple fields are
used, or slow roll is violated. Numerical simulations will be particularly
useful in analysing models with these characteristics.
We will discuss in
Section~\ref{sec:disc} planned future work to extend our current numerical
system to deal with these extensions beyond the standard single field slow roll
inflation.

In Section \ref{sec:perts} we will give a brief outline of
perturbation theory and describe the second order perturbation
equations that will be numerically calculated. Section
\ref{sec:numerics} describes the numerical implementation of the
calculation, including the initial conditions used and the
computational requirements. We present the results of this calculation
in Section \ref{sec:results} including a comparison of the second
order perturbation calculated for the $\msqphisq$ and
$\lambdaphifour$ potentials. 
We will discuss these results
and the next stages of this work in Section \ref{sec:disc}. 

Throughout this paper we set $\hbar=c=1$ and use the reduced
Planck mass $\Mpl=\sqrt{8\pi G}$. Overdots and primes denote
differentiation with respect to our time variable $n$ (the number of
e-foldings) and conformal time $\eta$, respectively, and will be
defined explicitly when first used.  We will work in a flat
Friedmann-Robertson-Walker (FRW) background.

\section{Perturbations}
\label{sec:perts}

In this section we will briefly review the derivation of first and
second order perturbations in the uniform curvature gauge and describe
the slow roll approximation that we will use in this paper. There are
many reviews on the subject of cosmological perturbation theory, and
here we will follow Ref.~\cite{Malik:2008im}.  The full closed
Klein-Gordon equation for second order perturbations was recently
given by one of the authors and we will outline the derivation in
Ref.~\cite{Malik:2006ir} below.

\subsection{First and Second Order}
\label{sec:foandsoperts}

In this paper we will consider perturbations of a single scalar field
and will work throughout in the uniform curvature or flat gauge. Our
goal is to describe scalar perturbations up to second order and the
first step to achieve this is to examine the metric tensor:
\begin{eqnarray}
\label{metric1}
g_{00}&=&-a^2\left(1+2\phi_1+\phi_2\right) \,, \\
g_{0i}&=&a^2\left(B_1+\frac{1}{2}B_2\right)_{,i}\,, \\
g_{ij}&=&a^2\left[\left(1-2\psi_1-\psi_2\right)\delta_{ij}
+2E_{1,ij}+E_{2,ij}\right]\,,
\end{eqnarray}
where $a=a(\eta)$ is the scale factor, $\eta$ conformal time,
$\delta_{ij}$ is the flat background metric, $\phi_1$ and $\phi_2$ the
lapse functions, and $\psi_1$ and $\psi_2$ the curvature perturbations
at first and second order; $B_1$ and $B_2$ and $E_1$ and $E_2$ are
scalar perturbations describing the shear.
Spatial 3-hypersurfaces are flat in our chosen gauge and so 
\begin{equation}
 \label{defgauge}
\wt\psi_1=\wt\psi_2=\wt E_1=\wt E_2=0 \,,
\end{equation}
where the tilde denotes quantities in flat gauge.

The Sasaki-Mukhanov variable, i.e.\ the field perturbation on uniform curvature
hypersurfaces \cite{Sasaki:1986hm,Mukhanov:1988jd}, evaluated at first order is
given by 
\begin{equation}
\label{defQ1I}
\wt{\dvp1}=\dvp1+\frac{\vp_{0}'}{\H}\psi_1\,, 
\end{equation}
where $\vp_0$ is the background value of the field and the perturbations of
$\varphi$ are defined as 
\begin{equation}
 \varphi(x^\mu) = \vp_0(\eta) + \dvp1(\eta, x^i) + \frac{1}{2}\dvp2(\eta, x^i)
\,.
\end{equation}
At second order the Sasaki-Mukhanov variable becomes more complicated
\cite{Malik:2005cy,Malik:2003mv}:
\begin{equation}
\label{defQ2I}
\wt{\dvp2}=\dvp2+\frac{\vp_{0}'}{\H}\psi_2
+\left(\frac{\psi_1}{\H}\right)^2\left[
2\H\vp_{0}'+\vp_{0}''-\frac{\H'}{\H}\vp_{0}'\right]
+2\frac{\vp_{0}'}{\H^2}\psi_1'\psi_1+\frac{2}{\H}\psi_1\dvp1'
-2\delta\vp_{1,k}E_{1,}^{~k}
+{\cal{X}}\left(\psi,E\right)\,,
\end{equation}
where ${\cal{X}}\left(\psi,E\right)$ contains terms quadratic in
gradients of the metric perturbations $\psi_1$ and $E_1$. From now on we will
drop the tildes and talk only about variables in the flat gauge.
The potential of the scalar field is also split
\begin{equation}
 U(\varphi) = \U + \delta U_{1} + \frac{1}{2}\delta U _{2}\,,\quad
 \delta U_{1} = \Uphi \dvp1 \,,\quad
 \delta U_{2} = \Upp \dvp1^2 + \Uphi\dvp2 \,,
\end{equation}
where $\Uphi = \pd{U}{\varphi}$.
The Klein-Gordon equation describes the evolution of the scalar field. For the
background field we have
\begin{equation}
\label{KGback}
\vp_{0}''+2\H\vp_{0}'+a^2 \Uphi = 0\,,   
\end{equation}
where $\H\equiv\frac{a'}{a}$ is related to the Hubble parameter $H$ by $\H=aH$.
The first order equation is
\begin{equation}
\label{KG1flatsingle}
\dvp1''+2\H\dvp1'+2a^2 \Uphi \phi_1
-\nabla^2\dvp1-\vp_{0}'\nabla^2 B_1
-\vp_{0}'\phi'_1 + a^2 \Upp \dvp1
=0\,,
\end{equation}
and the second order
\begin{eqnarray}
\label{KG2flatsingle}
\dvp2''&+&2\H\dvp2'-\nabla^2\dvp2+a^2 \Upp \dvp2
+a^2 \Uppp (\dvp1)^2 +2a^2 \Uphi \phi_2
-\vp_{0}'\left(\nabla^2 B_2+\phi_2'\right)\nonumber\\
&+&4\vp_{0}' B_{1,k}\phi_{1,}^{~k}
+2\left(2\H\vp_{0}'+a^2 \Uphi\right) B_{1,k}B_{1,}^{~k}
+4\phi_1\left(a^2 \Upp \dvp1-\nabla^2\dvp1\right)
+4\vp_{0}'\phi_1\phi_1'\nonumber\\
&-&2\dvp1'\left(\nabla^2 B_1+\phi_1'\right)-4\dvp1'_{,k}B_{1,}^{~k}
=0\,,
\end{eqnarray}
where as mentioned before all the variables are now in the flat gauge.

The Einstein field equations are also required at first and second order. We
will not reproduce them here but instead refer the interested reader to Section
II B of \Rref{Malik:2006ir}.
Using the perturbed Einstein equations, the Klein-Gordon equations above can be
written in closed form at both first and second orders. These equations will
form the basis of the numerical scheme described in Section \ref{sec:numerics}.

The dynamics of the scalar field becomes clearer in Fourier space but terms in
the second order equation of the form $\left(\dvp1(x)\right)^2$ require the use
of the convolution theorem (see for example \Rref{Vretblad:2005}).
Following Refs.~\cite{Malik:2006ir} and \cite{book:liddle} we will write $\dvp{}(k^i)$
for the Fourier component of $\dvp{}(x)$ such that
\begin{equation}
 \dvp{}(\eta, x^i) = \frac{1}{(2 \pi)^3} \int d^3k \dvp{}(\kvi) \exp (i k_i x^i)
\,,
\end{equation}
where $\kvi$ is the comoving wavenumber.

In Fourier space the closed form of the first order Klein-Gordon equation
transforms into
\begin{equation}
\label{eq:fokg}
 \dvp1(\kvi)'' + 2\H \dvp1(\kvi)' + k^2\dvp1(\kvi) + a^2 \left[\Upp +
\frac{8\pi G}{\H}\left(2\vp_{0}' \Uphi + (\vp_{0}')^2\frac{8\pi G
}{\H}\U\right)\right]\dvp1(\kvi) = 0 \,.
\end{equation}
As mentioned above the second order equation requires more careful
consideration with terms quadratic in the first order perturbation, which
 require convolutions of the form
\begin{equation}
 f(x)g(x) \longrightarrow \frac{1}{(2 \pi)^3} \int d^3q d^3p\, \delta^3(\kvi -\pvi -\qvi) f(\pvi)
g(q^i) \,.
\end{equation}
For convenience we will group the terms with gradients of $\dvp1(x)$
together and denote them by $F$. 
The full closed form second order Klein-Gordon
equation in Fourier Space is 
\begin{eqnarray}
\label{eq:SOKG_real}
 &&\dvp2''(\kvi) + 2\H \dvp2'(\kvi) + k^2 \dvp2(\kvi)
+ a^2\left[\Upp + \frac{8\pi G}{\H}\left(2\vp_{0}'\Uphi
+ (\vp_0')^2\frac{8\pi G}{\H}\U \right) \right]\dvp2(\kvi) \nonumber \\
 && + \frac{1}{(2\pi)^3}\int d^3q d^3p\, \delta^3(\kvi -\pvi -\qvi) \left\{ \frac{16\pi G}{\H}
\left[ X
\dvp1'(\pvi) \dvp1(\qvi) + \vp_{0}' a^2\Upp \dvp1(\pvi)\dvp1(\qvi)
\right] \right. \nonumber \\
 && + \left(\frac{8\pi G}{\H}\right)^2\vp_{0}'\left[2a^2\Uphi\vp_{0}'
\dvp1(\pvi)\dvp1(\qvi) + \vp_{0}'X\dvp1(\pvi)\dvp1(\qvi) \right]
\nonumber \\
 && -2\left(\frac{4\pi G}{\H}\right)^2\frac{\vp_{0}' X}{\H} \left[X\dvp1(\pvi) \dvp1(\qvi) +
\vp_{0}' \dvp1(\pvi) \dvp1'(\qvi)\right] \nonumber \\
 && \left. +\frac{4\pi G}{\H} \vp_{0}' \dvp1'(\pvi) \dvp1'(\qvi) 
 + a^2\left[\Uppp + \frac{8\pi G}{\H}\vp_{0}' \Upp\right] \dvp1(\pvi)
\dvp1(\qvi) \right\} \nonumber \\
 && + F(\dvp1(\kvi), \dvp1'(\kvi)) = 0\,.
\end{eqnarray}
Here we use $X=a^2 (8\pi G \U \vp_0'/\H + \Uphi)$ for convenience.
The $F$ term contains gradients of $\dvp1$ in real space and therefore
the convolution integrals include additional factors of $k$ and
$q$. It is given by
\begin{eqnarray}
 \label{Fdvk1_fourier}
&&
F\left(\dvp1(\kvi),\dvp1'(\kvi)\right)
= \frac{1}{(2\pi)^3}\int d^3pd^3q\delta^3(\kvi-\pvi-\qvi) 
\Bigg\{ 
2\left(\frac{8\pi G}{\H}\right)\frac{p_kq^k}{q^2}
\delta\vp_{1}'(\pvi)\left(
X\dvp1(\qvi)+\vp_{0}'\dvp1'(\qvi)\right)
\nonumber\\
&&
+p^2\frac{16\pi G}{\H}\dvp1(\pvi)\vp_{0}'\dvp1(\qvi)
+\left(\frac{4\pi G}{\H}\right)^2
\frac{\vp_{0}'}{\H}\Bigg[
\left(p_lq^l-\frac{p^iq_jk^jk_i}{k^2}\right) 
\vp_{0}'\delta\vp_{1}(\pvi)\vp_{0}'\delta\vp_{1}(\qvi)
\Bigg]\nonumber\\
&&
+2\frac{X}{\H}\left(\frac{4\pi G}{\H}\right)^2 
\frac{p_lq^lp_mq^m+p^2q^2}{k^2q^2}
\Bigg[\vp_{0}'\delta\vp_{1}(\pvi)
\left(X\dvp1(\qvi)+\vp_{0}'\dvp1'(\qvi)\right)
\Bigg]
\nonumber\\
&&
+\frac{4\pi G}{\H}
\Bigg[
4X\frac{q^2+p_lq^l}{k^2}\left(
\dvp1'(\pvi)\dvp1(\qvi)\right)
-\vp_{0}'p_lq^l \delta\vp_{1}(\pvi)\delta\vp_{1}(\qvi)
\Bigg]
\nonumber\\
&&
+\left(\frac{4\pi G}{\H}\right)^2
\frac{\vp_{0}'}{\H}\Bigg[
\frac{p_lq^lp_mq^m}{p^2q^2}
\left( X\dvp1(\pvi)+\vp_{0}'\dvp1'(\pvi)\right)
\left(X\dvp1(\qvi)+\vp_{0}'\dvp1'(\qvi)\right)
\Bigg]\nonumber\\
&&
+\frac{\vp_{0}'}{\H}
\Bigg[
8\pi G\left(\frac{p_lq^l+p^2}{k^2}q^2\dvp1(\pvi)\dvp1(\qvi)
-\frac{q^2+p_lq^l}{k^2}\dvp1'(\pvi)\dvp1'(\qvi)
\right)
\nonumber\\
&&\qquad\qquad\qquad
+\left(\frac{4\pi G}{\H}\right)^2
\frac{k^jk_i}{k^2}\Bigg(
2\frac{p^ip_j}{p^2}
\left(X\dvp1(\pvi)+\vp_{0}'\dvp1'(\pvi)\right)
X\dvp1(\qvi)
\Bigg)\Bigg]
\Bigg\}\,.
\end{eqnarray}

\subsection{Slow Roll approximation}
\label{sec:slowroll}

In order to establish the viability of a numerical calculation of the
Klein-Gordon equation we have confined ourselves in this paper to studying the
evolution in the slow roll approximation. In our case this involves taking
\begin{equation}
 \vp_{0}'' = \H \vp_{0}' \simeq 0\,,\quad
\frac{\left(\vp_{0}'\right)^2}{2a^2}\ll \U\,,
\end{equation}
such that $X=0$ and $\H^2 = (8\pi G/3) a^2 \U$. The slow roll parameter
$\epsilon_H$ as defined in Refs.~\cite{Malik:2006ir} and \cite{Seery:2005gb} 
(which is the square-root of the usual $\epsilon$) is given by
\begin{equation}
 \varepsilon_H = \sqrt{4\pi G} \frac{\vp_{0}'}{\H}\,.
\end{equation}
With this approximation the second order equation (\ref{eq:SOKG_real})
simplifies dramatically, and with the $F$ term included is
\begin{eqnarray}
 \label{KG2_fourier_sr}
&&\dvp2''(\kvi)+2\H\dvp2'(\kvi)+k^2\dvp2(\kvi)
+\left(a^2
\Upp-{24 \pi G}(\vp_{0}')^2\right)
\dvp2(\kvi) \\
&&+\int d^3p\ d^3q\ \delta^3(\kvi-\pvi-\qvi) \Bigg\{
a^2\left(\Uppp
+ \frac{8\pi G}{\H}\vp_{0}' \Upp\right)
 \dvp1(\pvi)\dvp1(\qvi)
+\frac{16\pi G}{\H}a^2
\vp_{0}'\Upp\dvp1(\pvi)\dvp1(\qvi)\Bigg\}
\nonumber \\
&&+ \frac{8\pi G}{\H}
\int d^3p\ d^3q\ \delta^3(\kvi-\pvi-\qvi)  \Bigg\{
\frac{8\pi G}{\H}\frac{p_l q^l}{q^2}\vp_{0}'\dvp1'(\pvi)
\dvp1'(\qvi)
+2p^2\vp_{0}' \dvp1(\pvi) \dvp1(\qvi)\nonumber\\
&&\qquad\qquad\qquad\qquad
+\vp_{0}'
\Bigg(
\left(\frac{p_lq^l+p^2}{k^2}q^2-\frac{p_lq^l}{2}\right)
\dvp1(\pvi)\dvp1(\qvi)
+\left(\frac{1}{2}-\frac{q^2+p_lq^l}{k^2}\right)
\dvp1'(\pvi)\dvp1'(\qvi)\Bigg)
\Bigg\}=0 \,.\nonumber
\end{eqnarray}
The numerical simulation in this paper will solve the slow roll
version of the second order above, \eq{KG2_fourier_sr}, the first
order equation (\ref{eq:fokg}) and the background equation
(\ref{KGback}). In the next section we set up the correct form of
these equations for the numerical simulation and discuss the
implementation and some tests of the accuracy of the method.

\section{Numerics}
\label{sec:numerics}

Our goal in this paper is to show that, just as at first order, a
direct numerical calculation of the second order perturbations of a
scalar field system is achievable and in this section we will outline
how we have implemented this system. In structuring the numerical
system we have closely followed the work done at first order by Martin
and Ringeval \cite{Martin:2006rs, Ringeval:2007am} and previously by
Salopek \etal \cite{Salopek:1988qh}.

A finite numerical range of $k$ modes to be calculated is required.
The upper cutoff in $k$, which marks the smallest scale considered, is well
motivated by the difficulty in observing primordial perturbations at these small
scales. 
At the other end we need to specify the largest scale or smallest $k$ that we
will consider. Analytically this is often taken to be the size of the universe,
with $k=0$ being the equivalent mode. One immediate problem with this is that
the Bunch-Davies vacuum initial conditions outlined in Section
\ref{sec:initconds} blow up. The standard workaround is to implement a cutoff
at large scales beyond which the amplitude of perturbations is zero. This is a
pragmatic approach but recently there has been some evidence that a sharp
cutoff similar to this could be responsible for the lack of power at large
scales in the WMAP data \cite{Lyth:2007jh, spergel, Sinha:2005mn,Kim:2009pf}.
 
The main concern is that the $k$ range covers most if not all the modes observed
to date in the CMB. The WMAP team rely for their main results,
\cite{Komatsu:2008hk},  
 on $\ell$ multipoles in the range $\ell \in [3, 1000]$
which corresponds approximately\footnotemark  to $k\in \left[0.92 \e{-60}, 3.1 \times
10^{-58}\right]\Mpl = \left[3.5\e{-4}, 0.12\right] \Mpc^{-1}$.
\footnotetext{The approximate conversion for $\ell$ is $\ell\simeq \frac{2k}{H_0}$ and a Megaparsec
is given in Planck units as $1\Mpc^{-1} \simeq 2.6247\e{-57} \Mpl$.}
We will consider a similar range of $k$ modes in this paper, taking three
different ranges outlined in Section \ref{sec:results}. The
choice of $k$ range is flexible with the only constraint being that the number
of modes at second order is one greater than a power of two. This enables faster
integration using the Romberg method as explained below.

\subsection{Equations}
\label{sec:numeqs}

The equations in Section \ref{sec:slowroll} are not set up for a
numerical calculation and in this section we rearrange them into a more
suitable form. This involves a change of time coordinate and grouping of terms
into smaller units for calculation.
The second order slow roll equation (\ref{KG2_fourier_sr}) can be further
simplified by performing the $p$
integral and changing to spherical polar coordinates $q, \theta, \omega$ where
$q=|\textbf{q}|$. The $d^3q$ integral becomes
\begin{equation}
 \int d^3q \longrightarrow \int_{0}^{\infty} q^2 dq \int_{0}^{\pi}\sin \theta
d\theta 
   \int_{0}^{2\pi}d\omega \,.
\end{equation}
For each $k$ mode equation we take the $\theta=0, \omega=0$ axis in the
direction of $\kvi$, so that the angle between $\kvi$ and $\qvi$ is
$\theta$ and the scalar product $q_i k^i = q k \cos\theta$. 
The argument of
each $\dvp1$ or $\dvp1'$ term depends on $\theta$ through
$|\kvi-\qvi|=\sqrt{k^2 + q^2 -2kq \cos\theta}$ and
so must remain inside the $\theta$ integral. There is no $\omega$ dependence
in $\dvp1$ with this choice of axes, so the last integral is simply evaluated.

In the slow roll case there are only four different $\theta$ dependent terms,
here labelled A--D:
\begin{eqnarray}
\label{AtoD}
 A(\kvi,\qvi) &=& \int_0^\pi \sin(\theta) \dvp1(\kvi-\qvi) d\theta \,,
\nonumber\\
 B(\kvi,\qvi) &=& \int_0^\pi \cos(\theta)\sin(\theta) \dvp1(\kvi-\qvi)
d\theta \,,\nonumber\\
 C(\kvi,\qvi) &=& \int_0^\pi \sin(\theta) \dvp1'(\kvi-\qvi) d\theta \,,
\nonumber\\
 D(\kvi,\qvi) &=& \int_0^\pi \cos(\theta) \sin(\theta) \dvp1'(\kvi-\qvi)
d\theta \,.
\end{eqnarray}
Written using the terms in \eqs{AtoD} the slow roll equation
(\ref{KG2_fourier_sr}) becomes:
\begin{eqnarray}
\label{KG2_fourier_sr_aterms}
&&\dvp2''(\kvi)+2\H\dvp2'(\kvi)+k^2\dvp2(\kvi)
+\left(a^2
\Upp-{24 \pi G}(\vp_{0}')^2\right)
\dvp2(\kvi)
+ S(\kvi) = 0 \,,\\
\label{KG2_src_sr_aterms}
&&S(\kvi) = \frac{1}{(2\pi)^2}\int dq\ \Bigg\{
a^2\Uppp q^2 \dvp1(\qvi) A(\kvi,\qvi) \nonumber\\
&& \qquad\qquad\qquad\qquad\qquad + \frac{8\pi G}{\H}\vp_{0}' \Bigg[ 
\left( 3a^2\Upp + \frac{7}{2}q^4 + 2k^2q^2\right) A(\kvi,\qvi)
-\left(\frac{9}{2} + \frac{q^2}{k^2}\right)kq^3 B(\kvi,\qvi)
\Bigg]\dvp1(\qvi) \nonumber\\
&&\qquad\qquad\qquad\qquad \qquad+ \frac{8\pi G}{\H}\vp_{0}' \Bigg[
-\frac{3}{2}q^2 C(\kvi,\qvi) + \left(2-\frac{q^2}{k^2}\right)kq D(\kvi,\qvi) 
\Bigg]\dvp1'(\qvi) \Bigg\} \,,
\end{eqnarray}
where $S(\kvi)$ is the source term which will be determined before the
second order system is run. The full set of equations which must be evolved are
then \eq{KGback} for the background, \eq{eq:fokg} for the first
order perturbations and \eqs{KG2_fourier_sr_aterms} and
(\ref{KG2_src_sr_aterms}) for the second order and source terms.

A more appropriate time variable for the numerical simulation is the
number of e-foldings, and hence we use 
\begin{equation}
\label{def_silly_n}
n=\log(a/a_{\mathrm{init}})\,,
\end{equation}
as our time variable instead of conformal time. Here,
$a_{\mathrm{init}}$ is the value of $a$ at the beginning of the
simulation. If $a$ is set to be $1$ today we can calculate
$a_{\mathrm{init}}$ once the background run is complete and the end
time of inflation is determined as in Section
\ref{sec:implementation}. We will use an overdot to denote
differentiation with respect to $n$.

The changes in derivatives required are as follows:
\begin{eqnarray}
 \pd{ }{\eta} &=& \frac{d n}{d \eta}\pd{}{n} = \H \pd{}{n} \,,\\
 \pd{ }{t} &=& \frac{d \eta}{d t} \frac{d n}{d \eta}\pd{}{n} = H \pd{}{n}\,,
\end{eqnarray}
where $\eta$ and $t$ are conformal and coordinate time respectively with $H =
\frac{da}{dt}/a$ and $\H = aH$.
As mentioned above the value
of $a$ at the end of inflation is calculated using the connection equation (see for example
Eq.~(3.19) in \Rref{book:liddle} or Eq.~(7) in \Rref{Peiris:2008be}) assuming that instantaneous
reheating occurs
at the end of
inflation. This gives approximately 65 e-foldings from the end of inflation until now. 
The background and first order equations written in terms of the new time
variable $n$ are
\begin{eqnarray}
&&\ddot{\vp_{0}} + \frac{\U}{H^2}\dot{\vp_{0}} + \frac{\Uphi}{H^2} = 0 \,,
\label{eq:bgntime}\\
&&\ddot{\dvp1} + \left(3 + \frac{\dot{H}}{H}\right)\dot{\dvp1} +
\left[\left(\frac{k}{aH}\right)^2 + \frac{\Upp}{H^2} + \frac{8\pi G}{H^2}
2\dot{\vp_{0}}\Uphi + \left(\frac{8\pi G}{H}\right)^2
\left(\dot{\vp_{0}}\right)^2\U \right]\dvp1 = 0\,. \label{eq:fontime}
\end{eqnarray}
The second order equation in terms of $n$ is
\begin{equation}
 \label{KG2_fourier_sr_ntime}
\ddot{\dvp2}(\kvi)+\left(3 + \frac{\dot{H}}{H}\right)
\dot{\dvp2}(\kvi)+ \left(\frac{k}{aH}\right)^2\dvp2(\kvi)
+\left(\frac{\Upp}{H^2}-{24 \pi G}(\dot{\vp_{0}})^2\right)
\dvp2(\kvi) +S(\kvi) = 0 \,,
\end{equation}
\begin{eqnarray}
\label{KG2_source_ntime}
S(\kvi) &=& \frac{1}{(2\pi)^2}\int dq\ \Bigg\{
\frac{\Uppp}{H^2} q^2 \dvp1(\qvi) A(\kvi,\qvi) \nonumber\\
&&+\, \frac{8\pi G}{(aH)^2}\dot{\vp_{0}} \Bigg[ 
\left( 3a^2\Upp q^2 + \frac{7}{2}q^4 + 2k^2q^2\right) A(\kvi,\qvi) 
-\left(\frac{9}{2} + \frac{q^2}{k^2}\right)kq^3 B(\kvi,\qvi)
\Bigg]\dvp1(\qvi) \nonumber\\
&&+\, 8\pi G \dot{\vp_{0}} \Bigg[
-\frac{3}{2}q^2 \tilde{C}(\kvi,\qvi) + \left(2-\frac{q^2}{k^2}\right)kq
\tilde{D}(\kvi,\qvi) 
\Bigg]\dot{\dvp1}(\qvi) \Bigg\}\,,
\end{eqnarray}
where 
\begin{eqnarray}
 \tilde{C}(\kvi,\qvi) &=& \frac{1}{aH} C(\kvi-\qvi) = \int_0^\pi \sin(\theta)
\dot{\dvp1}(\kvi-\qvi) d\theta \,,\nonumber \\
 \tilde{D}(\kvi,\qvi) &=& \frac{1}{aH} D(\kvi-\qvi) = \int_0^\pi \cos(\theta)
\sin(\theta) \dot{\dvp1}(\kvi-\qvi)
d\theta \,.
\end{eqnarray}

The argument of $\dvp1$ and $\dot{\dvp1}$ in the $A$--$\tilde{D}$ terms requires
special consideration. 
To compute the integrals, $\theta$ is sampled at 
\begin{equation}
\label{eq:nthetadefn}
N_\theta = 2^l + 1
\end{equation}
points in the range
$[0,\pi]$ (for some $l\in \mathbb{N}$ to allow Romberg integration) and the magnitude of $\kvi-\qvi$
is
found using
\begin{equation}
 |\kvi -\qvi| = \sqrt{k^2 + q^2 - 2kq\cos(\theta)}\,.
\end{equation}
While $\dvp1(\kvi)=\dvp1(k)$, the value of $|\kvi-\qvi|$ is at most
$2\kmax$ where $k,q \in [\kmin,\kmax]$. This means that to calculate
the source term for the $k$ range described we require that $\dvp1$
and $\dot{\dvp1}$ be known in the range $[0, 2\kmax]$. In
Section \ref{sec:implementation} we will 
show that this first order upper bound does not significantly affect
performance. On the other hand $|\kvi-\qvi|$ can also drop below the
lower cutoff of calculated $k$ modes. As discussed above we will implement a sharp cut off and
take $\dvp1(k)=0$ for the values below
$\kmin$. When $\Delta k \simeq \kmin$ this affects only the $k=q$ modes and
is only significant close to $\kmin$. Section \ref{sec:codetests}
describes how the accuracy is affected by changing $\Delta k$ and
other parameters. Without extrapolating outside our computed $k$ range
it appears to be very difficult to avoid taking this small number of
$\dvp1$s to be zero.

The value of $|\kvi-\qvi|$ will not in general coincide with the computed $k$
values of $\dvp1$. We use linear interpolation between the closest $k$s to
estimate the value of $\dvp1$ at these points. We leave to future work the
implementation of a more
accurate but also numerically intensive interpolation scheme.

Throughout the discussion above we have not specified any particular
potential $U$ and indeed the numerical code can use any reasonable
potential provided that it gives a period of inflationary expansion in
the e-folding range being simulated. In this paper we have used the two
standard potentials $U=\msqphisq$ and $U=\lambdaphifour$
but a modular system allows another potential to be used instead. We
choose the parameters $m$ and $\lambda$ in agreement with the first
order perturbation results from WMAP5 at the pivot scale
$\kwmap=0.002\Mpc^{-1} \simeq5.25\times10^{- 60} \Mpl$ with the values $m=6.3267\times10^{-6}
\Mpl, \lambda=1.5506\times10^{-13}$.

\subsection{Initial Conditions} 
\label{sec:initconds}

The background system requires initial conditions for $\vp_{0},
\dot{\vp_{0}}$ and $H$. These initial conditions and the range of
e-foldings to be simulated must be selected with the choice of
potential in mind. Not only must the
e-folding range include an inflationary period, but the $k$ modes to
be calculated at first and second order must begin inside the horizon
during this range. For example the initial conditions $\vp_0 = 18\Mpl,
\dot{\vp_0} = -1 \Mpl, H = 4.65\e{-5}\Mpl$ for the $\msqphisq$ model 
give the background evolution described below and shown in Figure~\ref{fig:eps}.

The initial conditions are set for each $k$ mode a few e-foldings
before horizon crossing. This follows the example of Salopek
\etal
\cite{Salopek:1988qh} and is justified on the basis that the mode is
sufficiently inside the
horizon for the Minkowski limit to be taken. This initial time,
$n_{\mathrm{init}}(k)$, is calculated to be when 
\begin{equation}
 \frac{k}{aH|_{\mathrm{init}}} = 50 \,.
\end{equation}
The range of e-foldings being used must include the starting point for
all $k$ modes, but the parameter on the right hand side, here chosen to
be 50, can be changed if needed.  We use the small wavelength solution
of the first order equations as the initial conditions \cite{Salopek:1988qh}, with
\begin{eqnarray}
\label{eq:foics}
 \dvp1|_{\mathrm{init}} &=& \frac{\sqrt{8\pi G}}{a}
\frac{e^{-i k\eta}}{\sqrt{2k}} \,,\\
 \dot{\dvp1}|_{\mathrm{init}} &=& -\frac{\sqrt{8\pi G}}{a}
\frac{e^{-i k\eta}}{\sqrt{2k}} \left(1 + i \frac{k}{a H}\right) \,,
\end{eqnarray}
where the conformal time $\eta$ can be calculated from $\eta=\int dn/aH \simeq
-(aH(1-\epsilon_H))^{-1}$, when $\epsilon_H$ changes slowly. For example $\kwmap$ is initialised
about $65$ e-foldings before the end of inflation and crosses the horizon about $5$ e-foldings
later.
We also use these formulae in the calculation of the source term in \eq{KG2_source_ntime} to
determine the value of $\dvp1$ for a $k$ mode before its evolution starts.

We are interested in the production of second order effects by the
evolution of the the gaussian first order modes and we make no
assumptions about the existence of second order perturbations before
the simulation begins. Therefore we set the initial condition for each second order
perturbation mode to be $\dvp2=0, \dot{\dvp2}=0$ at
the time when the corresponding first order perturbation is initialised.

\subsection{Implementation} 
\label{sec:implementation}

\begin{figure}
\centering
 \includegraphics[scale=0.8]{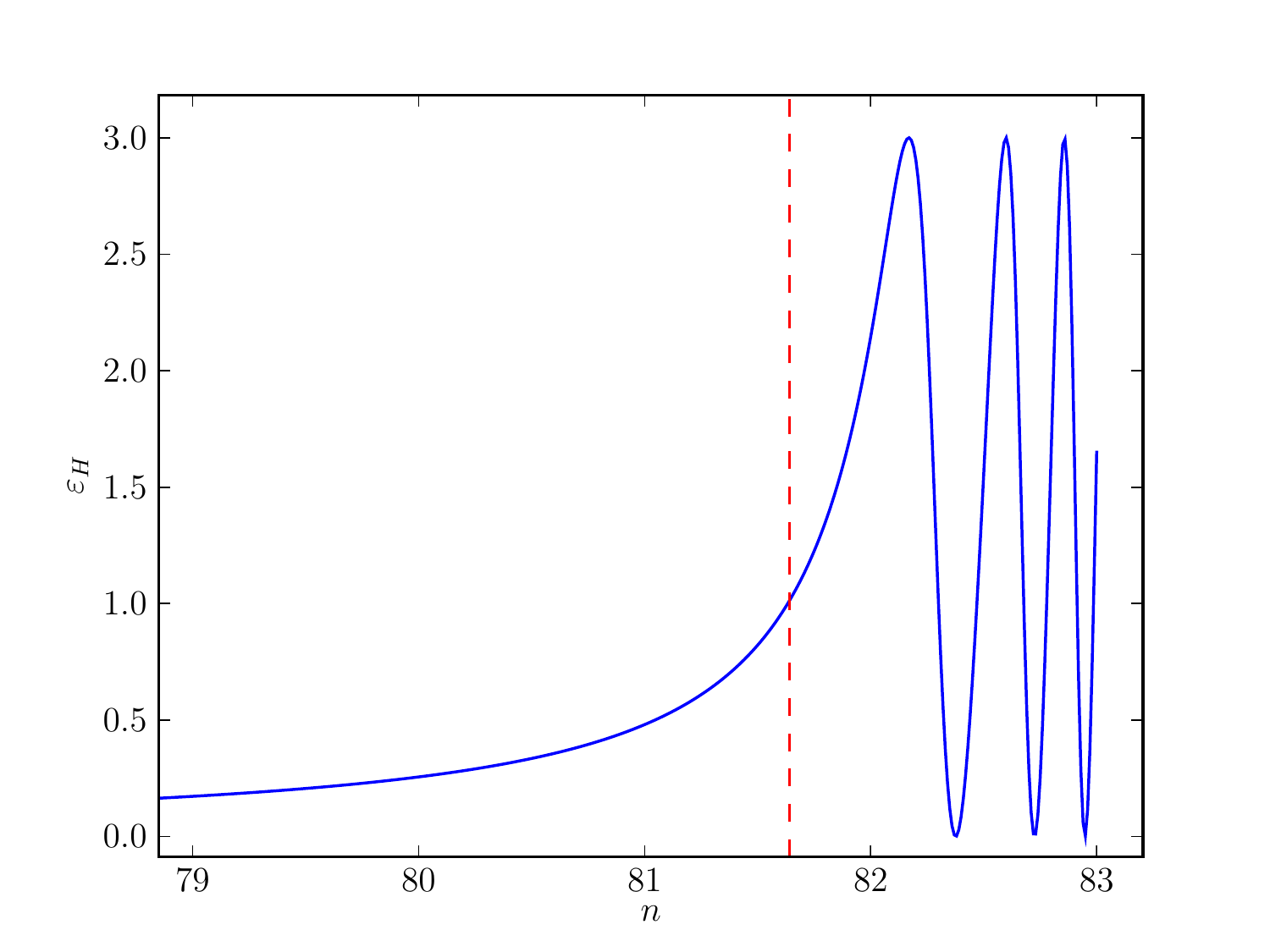}
 \caption{The end of inflation is determined by calculating when
   $\varepsilon_H=-\dot{H}/H=1$ (red dashed line). Along the $x$-axis,
   $n$ is the number of e-foldings from the start of the
   simulation.}
\label{fig:eps}
\end{figure}

The current implementation of the code is mainly in Python and uses the
Numerical and Scientific Python modules for their strong compiled array support
\cite{scipy}. The core of the model computation is a customised
Runge-Kutta 4th order method (see for example Eq.~(25.5.10) in
\cite{abramowitz+stegun}).  Following
Refs.~\cite{Martin:2006rs,Ringeval:2007am} the numerical calculation
proceeds in four stages. The background equation (\ref{eq:bgntime}),
rewritten as two first order (in the time derivative) equations, is
evolved from the specified initial state until some end time required
to be after the end of the inflationary regime.  The end of inflation
occurs when $d^2a/dt^2$ is no longer positive and the parameter
$\varepsilon_H = -\dot{H}/H$ first becomes greater than or equal to $1$
(see Figure~\ref{fig:eps}). Here, this specifies a new end time for the 1st
order run, although the simulation can run beyond the strict end of
inflation if required. The initial conditions for the first order
system are then set as outlined above.

The system of ordinary differential equations for the first order
perturbations from \eq{eq:fontime} is calculated using a standard
Runge-Kutta method. A fixed time step method is used in order to
simplify the construction of the  second order source term and because
\emph{a priori} it is not known which time steps would be required at
second order if an adaptive time step system were used. The first
order equations are separable in terms of $k$ and so it is
straightforward to run multiple instances of the system and collate
the results at the end. However, as will be discussed below, the first
order calculation is not computationally expensive in comparison with
the other stages and takes of the order of a few minutes for around
$8000$ time steps and $1025$ $k$ modes.

Once the first order system has been solved 
the source term for the second order system must be calculated. As the
real space equation for the source involves terms quadratic in the
first order perturbation it is necessary to perform a convolution in
Fourier space, as shown in \eq{KG2_fourier_sr_aterms}.  Transforming
back into real space was not considered due to the presence of both
gradient operators and their inverses. Here the slow roll version of
the source term integrand has been used, but the method can equally be
applied to the full equation. This stage is the most computationally
intensive and can be run in parallel as each time step is independent
of the others. The nature of the convolution integral and the
dependence of the first order perturbation on the absolute value of
its arguments requires that twice as many $k$ modes are calculated at
first order than are desired at second order as explained above.  As
the first order calculation is computationally cheaper than the source
term integration, this does not significantly lower the possible
resolution in $k$-space, which is still limited by the source term
computation time.  Once the integrand is determined it is fed into a
Romberg integration scheme. As for $\theta$  which was
discretised by $N_\theta$ points in \eq{AtoD}, this requires that the
number of $k$ modes is
\begin{equation}
\label{nk_constraint}
N_k=2^l + 1\,,
\end{equation}
for some\footnotemark $l\in\mathbb{N}$. 
\footnotetext{The number of discretised $k$ modes $N_k$ does not need to be equal to
  $N_\theta$.}
This requirement can be lifted by opting for a less
accurate and somewhat slower standard quadrature routine.

The second order system is finally run with the source term and other
necessary data being read as required from the memory or disk. The
Runge-Kutta method calculates half time steps for each required point,
for example if $y(x_n)$ is known and $y(x_{n+1})=y(x_n+h)$ is required
(for step size $h$), the method will calculate the derivatives of $y$
at $y(x_n), y(x_n +h/2)$ and $y(x_n + h)$. As we need to specify the
source term at every calculated timestep, the requested timestep for
the second order method must be twice that used at first order.  This decreases the
accuracy of the method but does not require the use
of splines and interpolation techniques to determine background and
1st order variables between time steps.

The second order system is similar in run time to the first order 
system but the
source integration is more complex, involving the
integration of $N_k^2\times N_\theta$ values at
each time step.
Although a large amount of data is produced at each step at this
stage for 
each of the wavenumbers $k$, only the integrated result is stored to be used
in the second order run.
Results for each stage are stored in the open HDF5 standard which can deal
efficiently with large
files and is very portable, allowing data analysis independent of the
Python/Numpy programming environment.
We intend to release the program under a suitable license once the code has
matured and some of the improvements discussed in Section \ref{sec:disc} have
been implemented.

\subsection{Code Tests}
\label{sec:codetests}

We have tested the numerical code in a variety of controlled
circumstances in order to quantify the effect of different choices of
parameters. In particular it is important to know whether the values
picked for $N_\theta$, the number of discretised $\theta$s, $\Delta k$, the size of the
spacing of the discretised $k$ modes, and the range of
$k$ values significantly impacts on the results. The ODE solving parts
of the code are straightforward and follow standard algorithms.

As mentioned above the WMAP results \cite{Komatsu:2008hk} use
observations in the range $k\in [0.92 \e{-60}, 3.1 \times
  10^{-58}]\Mpl = [3.5\e{-4}, 0.12] M_{\mathrm{pc}}^{-1}$. We will
consider three different $k$ ranges both in our results and the tests
of the code\footnotemark:
\begin{eqnarray}
\label{eq:Krangedefns}
K_1 &=& \left[1.9\e{-5}, 0.039\right]\Mpc^{-1}\,,\quad \Delta k = 3.8\e{-5}\Mpc^{-1} \,,\nonumber\\
K_2 &=& \left[5.71\e{-5}, 0.12\right]\Mpc^{-1}\,, \quad \Delta k = 1.2\e{-4}\Mpc^{-1}\,,
\nonumber\\ 
K_3 &=& \left[0.52\e{-5}, 0.39\right]\Mpc^{-1}\,, \quad \Delta k = 3.8\e{-4}\Mpc^{-1} \,.
\end{eqnarray}
\footnotetext{The $k$ ranges in $\Mpl$ are:
\begin{eqnarray*}
\label{eq:Krangedefns-mpl}
K_1 &=& \left[0.5\e{-61}, 1.0245\e{-58}\right]\Mpl\,,\quad \Delta k = 1\e{-61}\Mpl \nonumber\\
K_2 &=& \left[1.5\times 10^{-61}, 3.0735\e{-58}\right]\Mpl\,, \quad \Delta k = 3\e{-61}\Mpl
\nonumber\\ 
K_3 &=& \left[0.25\e{-60}, 1.02425\e{-57}\right]\Mpl\,, \quad \Delta k = 1\e{-60}\Mpl \,.
\end{eqnarray*}
}
The first, $K_1$, has a very fine resolution but covers only a small portion of the WMAP range. 
The next, $K_2$, is closest to the WMAP range with a still quite fine resolution.  The final
range, $K_3$, has a larger $k$ mode step size $\Delta k = 1\e{-60}\Mpl = 3.8\e{-4}\Mpc^{-1}$ and
covers a greater range than the others, extending to much smaller scales than WMAP can observe.

\begin{figure}
 \subfigure[The relative error for different $N_\theta$, the number of discretised $\theta$s,
 keeping the other
parameters fixed and using the $K_3$ range. The upper blue line ($N_\theta=129$)
and middle green line
 ($N_\theta=257$) have relative errors at least an order of magnitude larger than the lower red line
($N_\theta=513$).]{
 \includegraphics[scale=0.8]{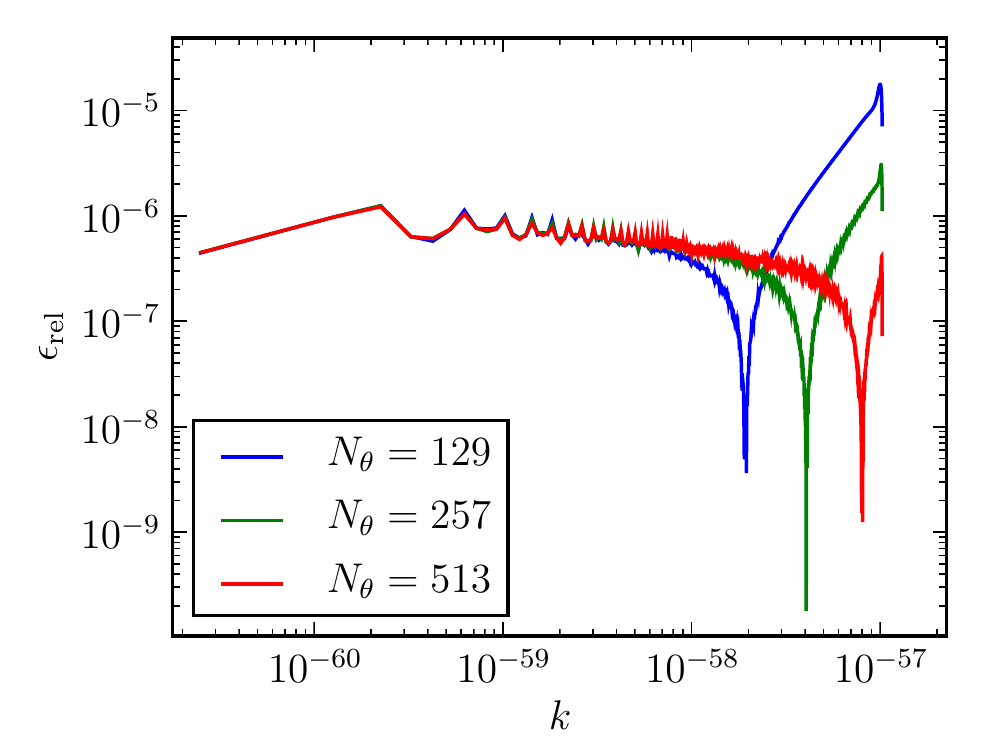}
 \label{fig:err_nthetas}
}
\subfigure[The relative error for the 3 different $k$ ranges $K1$, $K2$,
$K3$ (starting from the left). The parameter $\Delta k$
is set equal to $1\e{-61}\Mpl, 3\e{-61}\Mpl, 1\e{-60}\Mpl$ respectively.]{
 \includegraphics[scale=0.8]{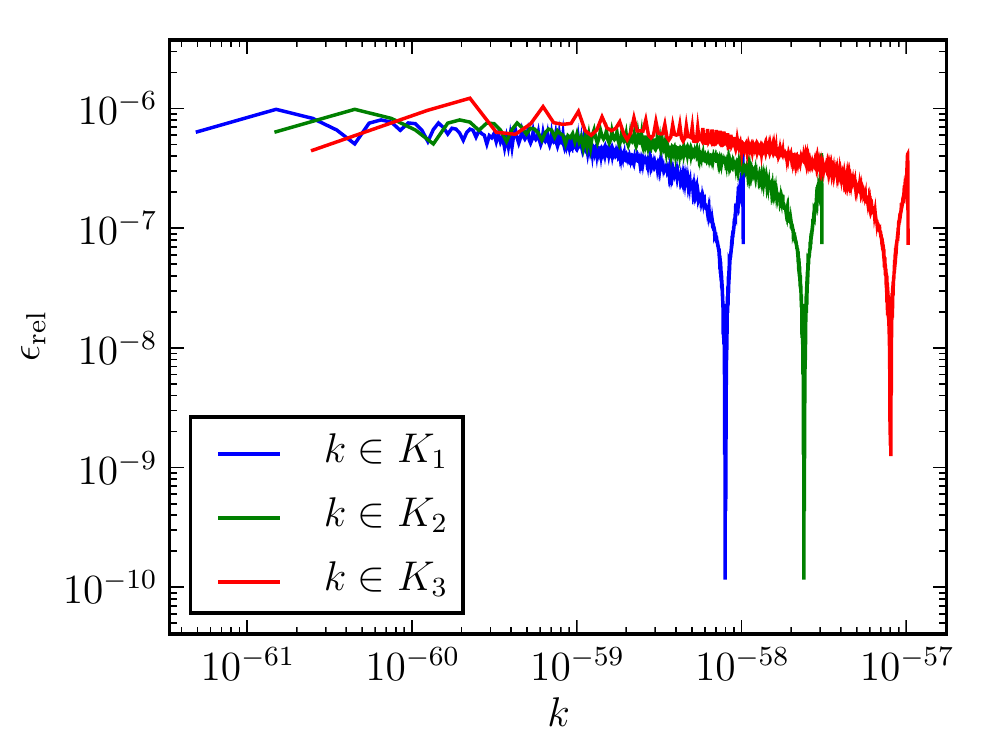}
 \label{fig:err-kranges}
}
\caption{Comparison of relative errors for different $N_\theta$ and $k$ ranges.}
\label{fig:err-comparison}
\end{figure}

The main new addition in the code is the calculation of the
convolution of the perturbations for the source term
\eq{KG2_source_ntime}. In particular the first of the $\theta$ dependent terms in \eq{AtoD}, $A$,
can be convolved analytically for certain smooth $\dvp1(k)$s. 
 We take $\dvp1(k)$ to be similar in form to the initial conditions
(\ref{eq:foics}), for example $\dvp1(k)\propto 1/\sqrt{k}$ with proportionality constant
$\alpha$.
If $I_A$ denotes the convolution of the $A$ term:
\begin{equation}
 I_A (k) = 2\pi \int_{\kmin}^{\kmax} q^2 \dvp1(q) A(k, q) dq \,,
\end{equation}
then putting in $\dvp1(k) = \alpha/\sqrt{k}$ gives
\begin{equation}
 I_A(k) = 2\pi \alpha^2 \int_{\kmin}^{\kmax} dq\, q^{\frac{3}{2}}
\int_{0}^{\pi} d\theta\, (k^2 + q^2 -2k q \cos{\theta})^{-4} \sin{\theta}\,. 
\end{equation}
This has the analytic solution
\begin{eqnarray}
\label{eq:err_analytic}
 A(k) &=& \frac{\pi}{18}\frac{\alpha^2}{k} \left\{ 
3k^3\left[\log\left(2\sqrt{k}\right) - \frac{\pi}{2}  + 
\arctan\left(\sqrt{\frac{\kmin}{k-\kmin}}\right) 
 + \log\left(2\left(\sqrt{\kmin} + \sqrt{\kmin + k}\right)\right) \right.\right. \nonumber \\
&& \left. -\, \log\left(2\left(\sqrt{\kmax} + \sqrt{\kmax + k}\right)\right) 
  - \log\left(2\left(\sqrt{\kmax} + \sqrt{\kmax - k}\right)\right) \right]  \nonumber \\
&& +\, \sqrt{\kmax}\left[\sqrt{\kmax -k} \left(-3k^2 + 14k\kmax - 8\kmax^2\right)  
 + \sqrt{\kmax +k} \left(3k^2 + 14k\kmax + 8\kmax^2\right)\right] \nonumber \\ 
&& \left. -\, \sqrt{\kmin}\left[\sqrt{k -\kmin} \left(3k^2 - 14k\kmax + 8\kmax^2\right) 
 - \sqrt{k +\kmin} \left(3k^2 + 14k\kmax + 8\kmax^2\right)\right] \right\} \,.
\end{eqnarray}
We have tested our code against this analytic solution for various
combinations of $k$ ranges and $N_\theta$. The relative error
\begin{equation}
 \epsilon_\mathrm{rel} = \frac{|\mathrm{analytic}- \mathrm{calculated} |}{|\mathrm{analytic}|}
\end{equation}
is small for all the tested cases but certain combinations of
parameters turned out to be better than others. The relative error of
all the following results is not affected by the choice of $\alpha$ so
we will keep it constant throughout.

\begin{figure}
 \centering
 \includegraphics[scale=0.8]{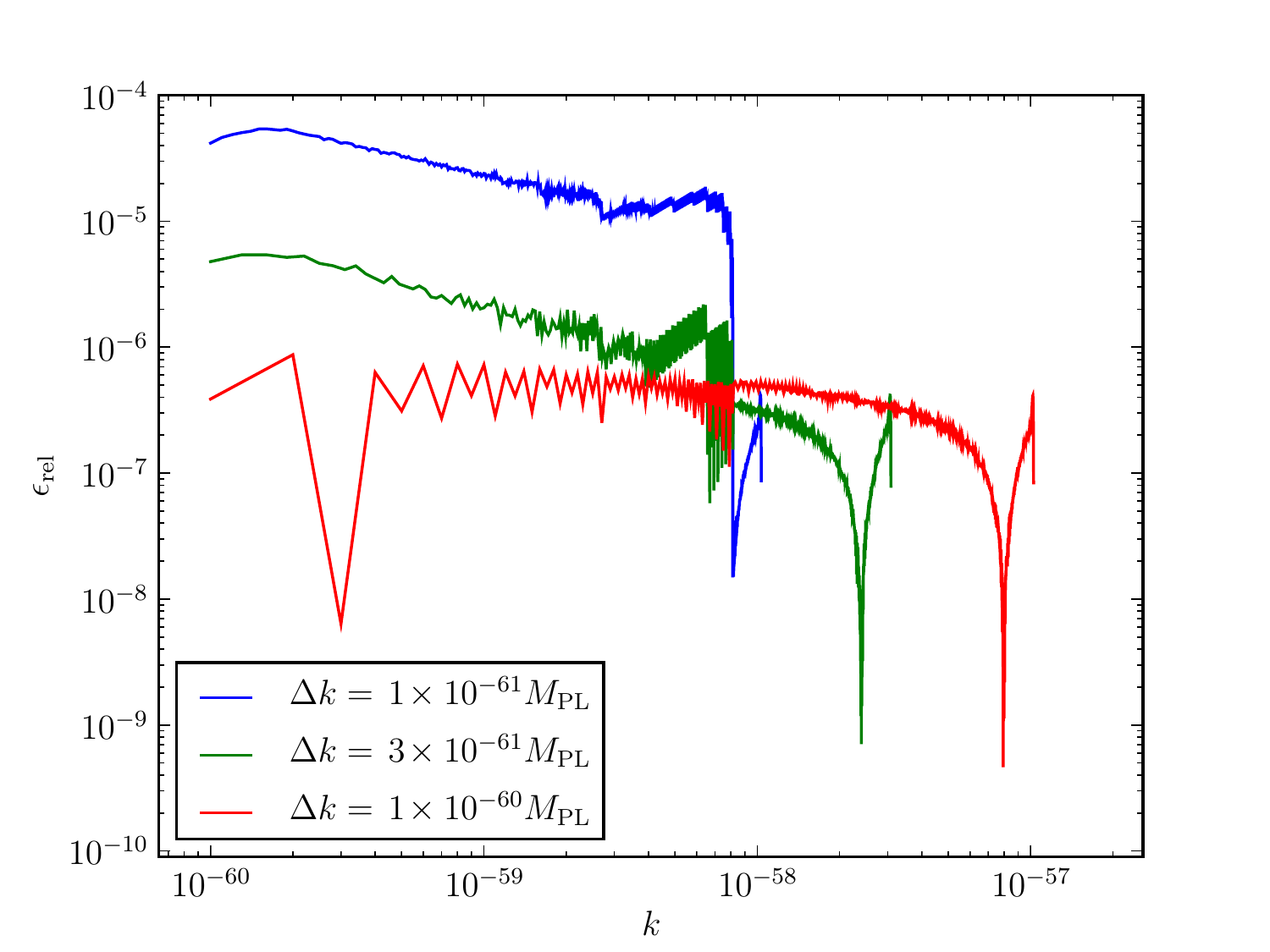}
 \caption{The relative error in convolution term $A$ for different values of $\Delta k$.
The other parameters are fixed: $\kmin=1\times10^{-60}\Mpl, N_k = 1025$ and $N_\theta=513$. The
upper 
blue line ($\Delta k=1\e{-61}\Mpl$ and middle green line ($\Delta k=3\e{-61}\Mpl$) have relative
errors at
least an order of magnitude larger than the lower red line ($\Delta k=1\e{-60}\Mpl$).}
 \label{fig:err-deltaks}
\end{figure}

We first tested the effect of changing $N_\theta$, the number of
samples of the $\theta$ range $[0,\pi]$.  Figure~\ref{fig:err_nthetas}
plots these results for the $k$ range $K_3$ with $\Delta k =
1\e{-60}\Mpl$. Only three values of $N_\theta$ are shown for clarity. It
can be seen that increasing $N_\theta$ decreases the relative error (for the convolution
term at least) when the other parameters are kept constant, as one
might expect.

As mentioned above the choice of $k$ range is especially important as
the convolution of the terms depends heavily on the minimum and
maximum values of this range. Indeed this is clear from the analytic
solution in \eq{eq:err_analytic}. Figure~\ref{fig:err-kranges}
shows the difference in relative error for the three different $k$
ranges described above with 
$\Delta k= 3.8\e{-5}, 1.2\e{-4}$ and $3.8\e{-4}\Mpc^{-1}$
($\Delta k= 1\e{-61}, 3\e{-61},1\e{-60}\Mpl$)
respectively. The accuracy is similar in all three cases.

Another important check is whether the resolution of the $k$ range is
fine enough. Varying $\Delta k$ can not be done in isolation of
course, if the constraint for $N_k$, \eq{nk_constraint},
is to
be met. For this test the end of the $k$ range changed with $\Delta k$
but the other parameters were kept fixed as $\kmin=1\e{-60}\Mpl=3.8\e{-4}\Mpc^{-1},
N_k = 1025$ and $N_\theta=513$. Figure~\ref{fig:err-deltaks} plots
these results again for only three indicative values.  For $\Delta
k<\kmin$, here the upper two lines, there is a marked degradation in
the accuracy of the method. This is understandable as many
interpolations of multiples of $\Delta k$ below $\kmin$ will be set to
$0$. Once $\Delta k$ is greater than $\kmin$ the relative error is
very similar for higher values (not shown in the figure).

\begin{figure}
 \centering
 \includegraphics[scale=0.8]{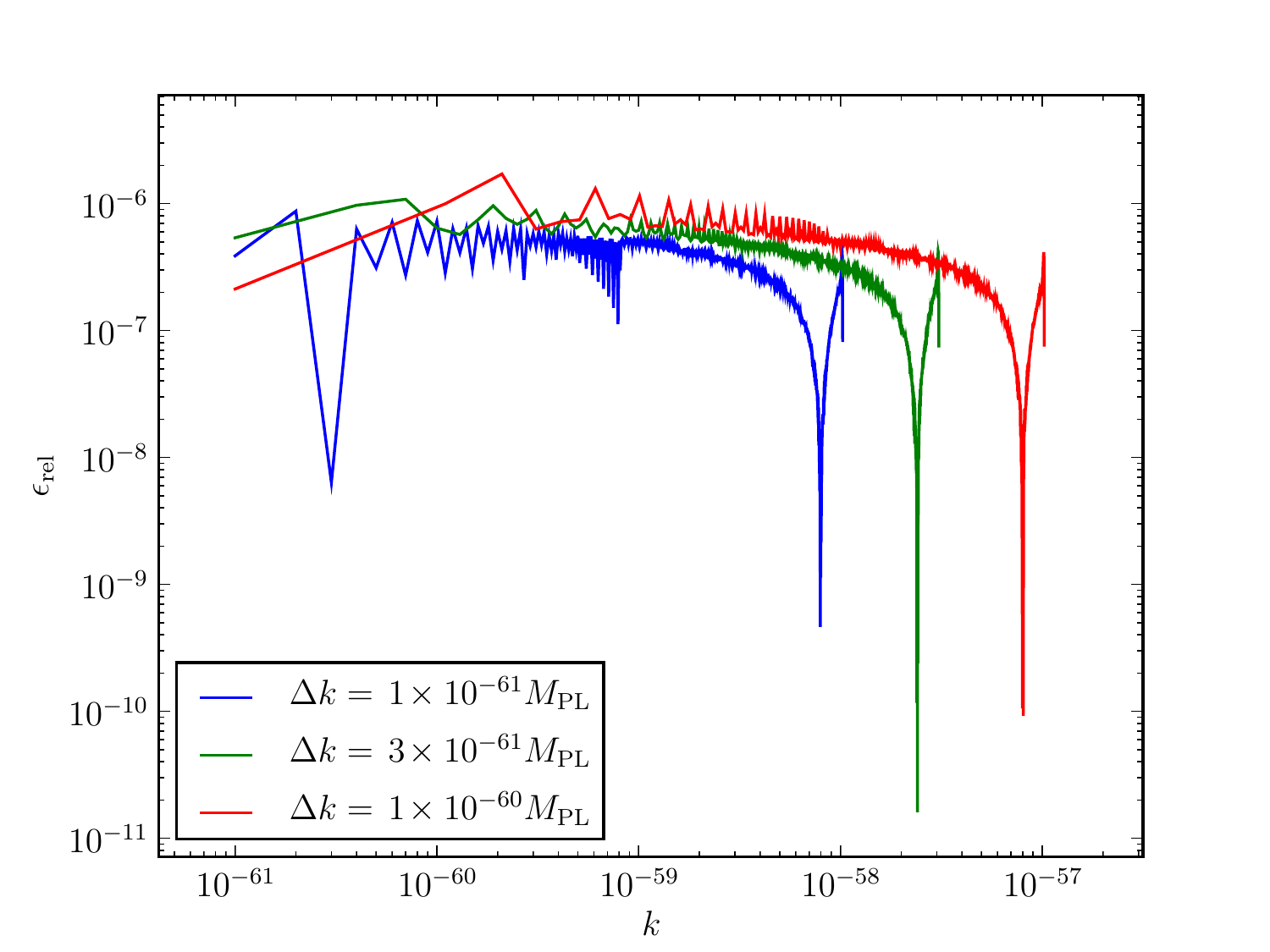}
 \caption{The relative error of the convolution term for three different values of $\Delta k$. In
contrast to Figure~\ref{fig:err-deltaks} $\kmin=1\e{-61}\Mpl=3.8\e{-5}\Mpc^{-1} \le \Delta k$ for
each.}
 \label{fig:err-deltak-kmin}
\end{figure}

It should be noted that these tests show the relative errors in the
computation of the $A$ convolution term, the most straightforward term in \eq{AtoD}, only and do not
represent
errors for the full calculation. However, they show that at least for
the pure convolution term the accuracy is good compared with the
analytic results. Equation (\ref{eq:err_analytic}) gives some indication of
the difficulty involved in finding an analytic solution for the other
terms, although this is a goal for future work. Having described the
implementation and accuracy of the numerical system we will outline
our results in the next section.

\section{Results}
\label{sec:results}

The main result of this paper is the demonstration of a numerical solution to
the closed Klein-Gordon equation of motion for second order scalar field
perturbations as described in \eq{KG2_fourier_sr}. This includes the slow
roll approximation of the source term for second order perturbations, but we have not used a slow
roll version of the evolution equations for the background or first order
perturbations. 

As a proof of concept we have tested the system with two standard potentials,
$\msqphisq$ and $\lambdaphifour$ and computed results across three
different $k$ ranges. As expected, considering the use of a single slowly
rolling field, the second order perturbation we have calculated is extremely
small in comparison with the first order term. However there are already
differences apparent between the two potentials which will be outlined below.
We have calibrated the parameters $m$ and $\lambda$ of the potentials using the
WMAP 5 normalisation at $\kwmap=0.002 \Mpc^{-1} = 5.25 \e{-60}\Mpl$
\cite{Komatsu:2008hk}.
We have outlined in \eq{eq:Krangedefns} the three $k$ ranges that we will use,
\begin{eqnarray*}
K_1 &=& \left[1.9\e{-5}, 0.039\right]\Mpc^{-1}\,,\quad \Delta k = 3.8\e{-5}\Mpc^{-1} \nonumber\\
K_2 &=& \left[5.71\e{-5}, 0.12\right]\Mpc^{-1}\,, \quad \Delta k = 1.2\e{-4}\Mpc^{-1}
\nonumber\\ 
K_3 &=& \left[0.52\e{-5}, 0.39\right]\Mpc^{-1}\,, \quad \Delta k = 3.8\e{-4}\Mpc^{-1} \,.
\end{eqnarray*}
Many of the results will be quoted for $\kwmap$ which lies in all three of these ranges.

Given that the first order perturbations for the chosen potentials give an
almost scale invariant power spectrum with no running, it is no surprise that
the results from the three different $k$ ranges are very similar. The second
order source term is somewhat dependent on the lower bound of $k$ (upper bound
on size). This is expected and in the scale invariant case a log divergence can
be shown to exist \cite{Lyth:2007jh}. We have implemented an arbitrary sharp cutoff at $\kmin$ below which 
$\dvp1$ is taken to be zero. As mentioned above there is some evidence to suggest that a 
similar cutoff is supported by the WMAP data \cite{Sinha:2005mn,Kim:2009pf}. 
%

\begin{figure}
 \centering
 \includegraphics[scale=0.8]{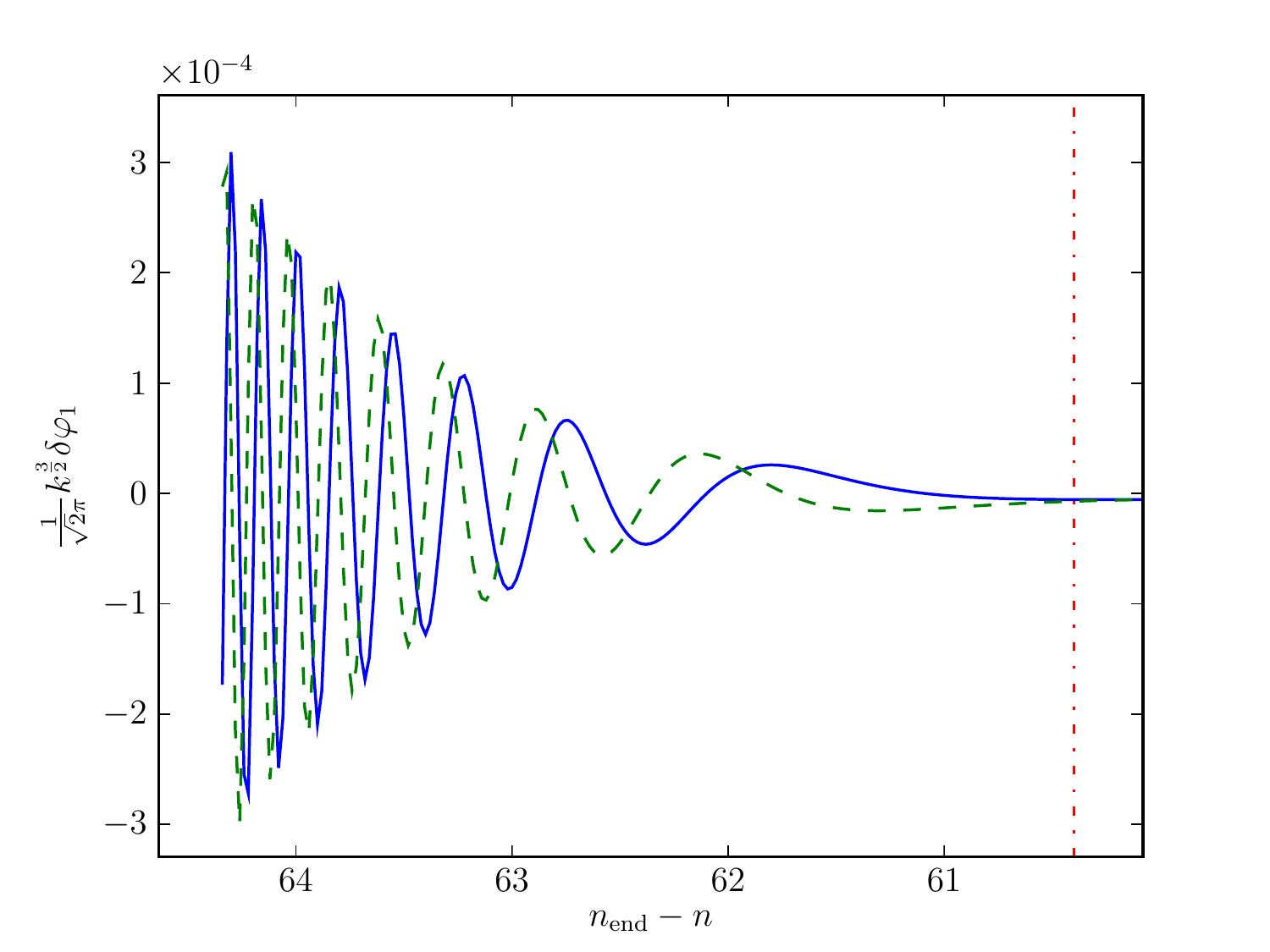}
 \caption{The first order perturbation $\dvp1$ rescaled by
$k^{3/2}/(\sqrt{2}\pi)$ from the beginning of the simulation until around
horizon crossing (red dot-dashed line). The real (blue) and imaginary (green
dashed) perturbations are shown for $\kwmap$.}
\label{fig:dp1}
\end{figure}

At first order our solutions agree with previous work
\cite{Salopek:1988qh,Martin:2006rs,Ringeval:2007am}, with oscillations
being damped until horizon crossing (when $k=aH$) after which the
curvature perturbation becomes conserved. Figure~\ref{fig:dp1} shows
the real and imaginary parts of the first order perturbations from
when the initial conditions are set at $k/aH=50$ to just after horizon
crossing. The x-axis for most of the following figures shows the
number of e-foldings left until the end of inflation instead of the
internally used time variable $n$.
%

\begin{figure}
 \centering
 \includegraphics[scale=0.8]{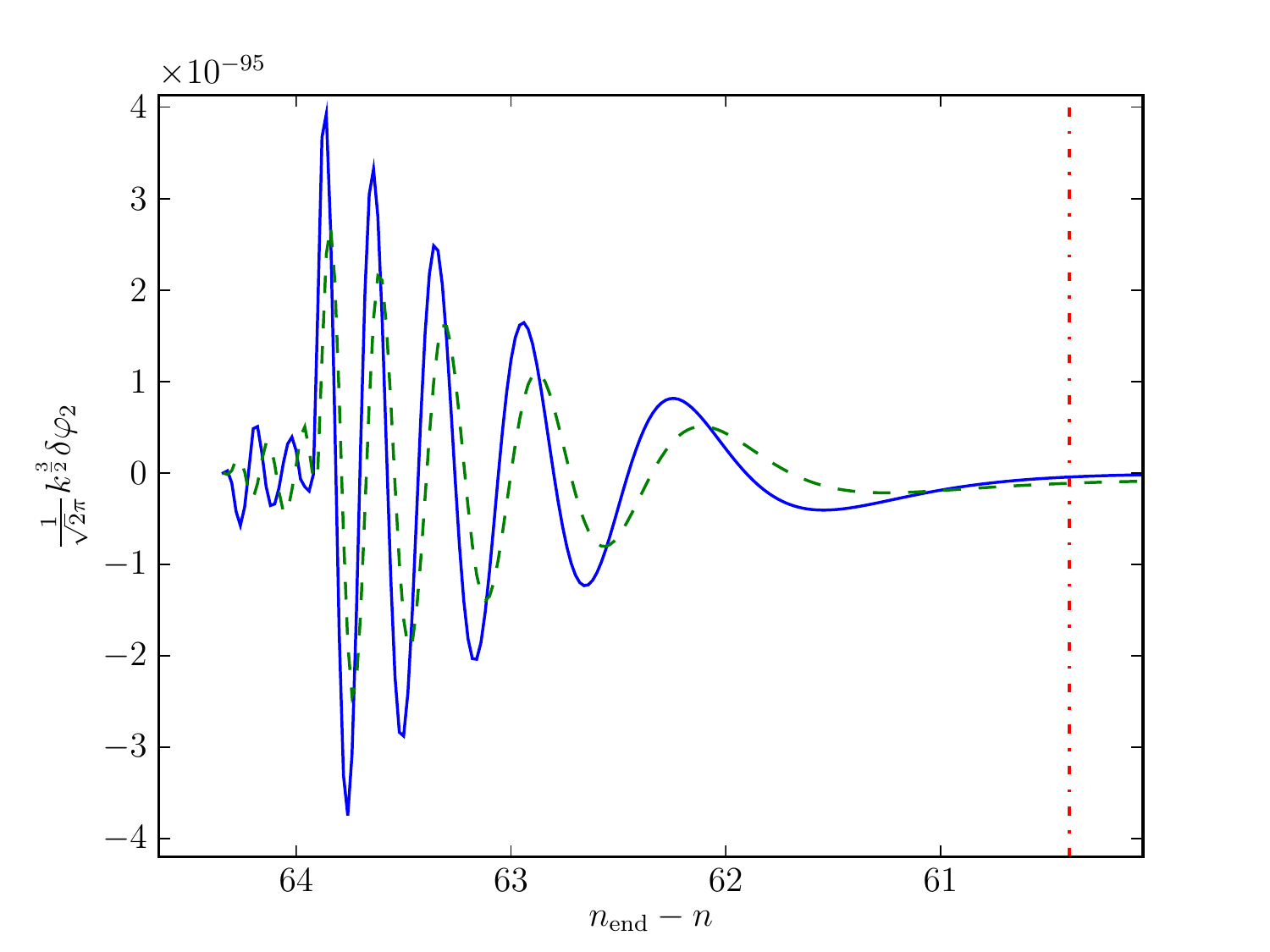}
 \caption{The real (blue line) and imaginary (green dashed) components of the second order
perturbation $\dvp2(\kwmap)$ from the beginning of the simulation until around the time
of horizon exit (red dot-dashed line).}
\label{fig:dp2realimag}
\end{figure}

In Figure~\ref{fig:dp2realimag} we show the evolution of the second
order perturbations for wavenumber $\kwmap$. As mentioned above the
overall amplitude of the second order perturbations is many orders of
magnitude smaller than the first order ones. In Figures~\ref{fig:dp1}
and \ref{fig:dp2realimag} the field values have been rescaled by
$k^{3/2}/(\sqrt{2}\pi)$ to allow a better appreciation of the
magnitude of the resulting power spectra.
%

\begin{figure}
\subfigure[Absolute magnitude of the source term.]{
 \includegraphics[scale=0.8]{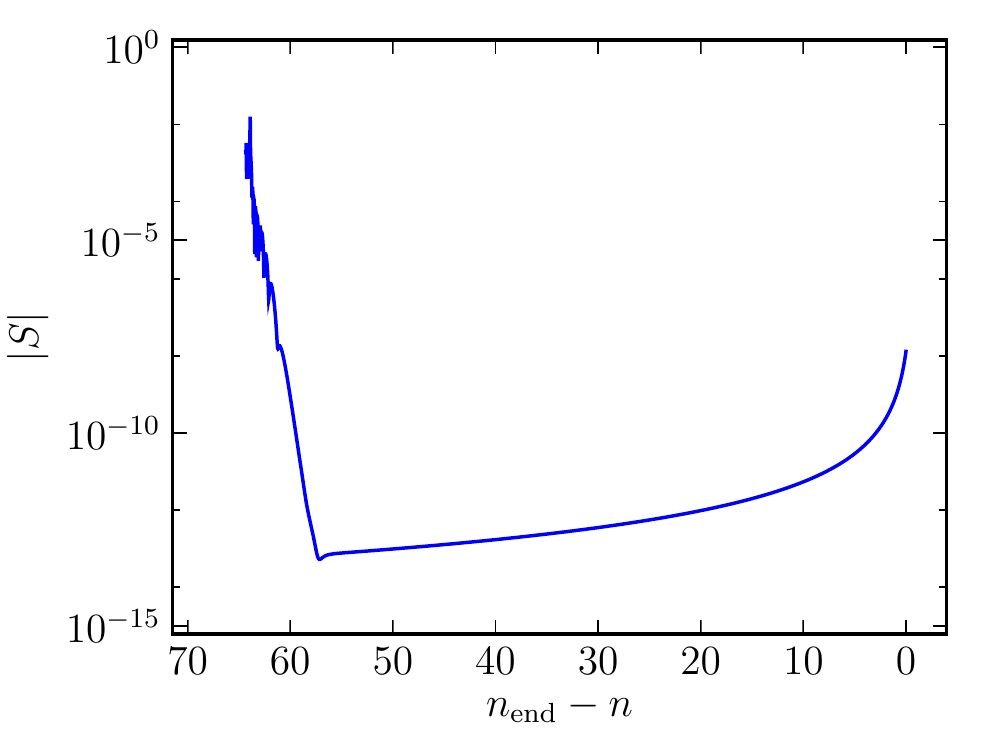}
\label{fig:src-full}
}
\subfigure[Power spectrum of scalar perturbations $\mathcal{P}_{\delta\varphi} = \frac{k^3}{2\pi^2}
|\delta\varphi|^2$.]{
 \includegraphics[scale=0.8]{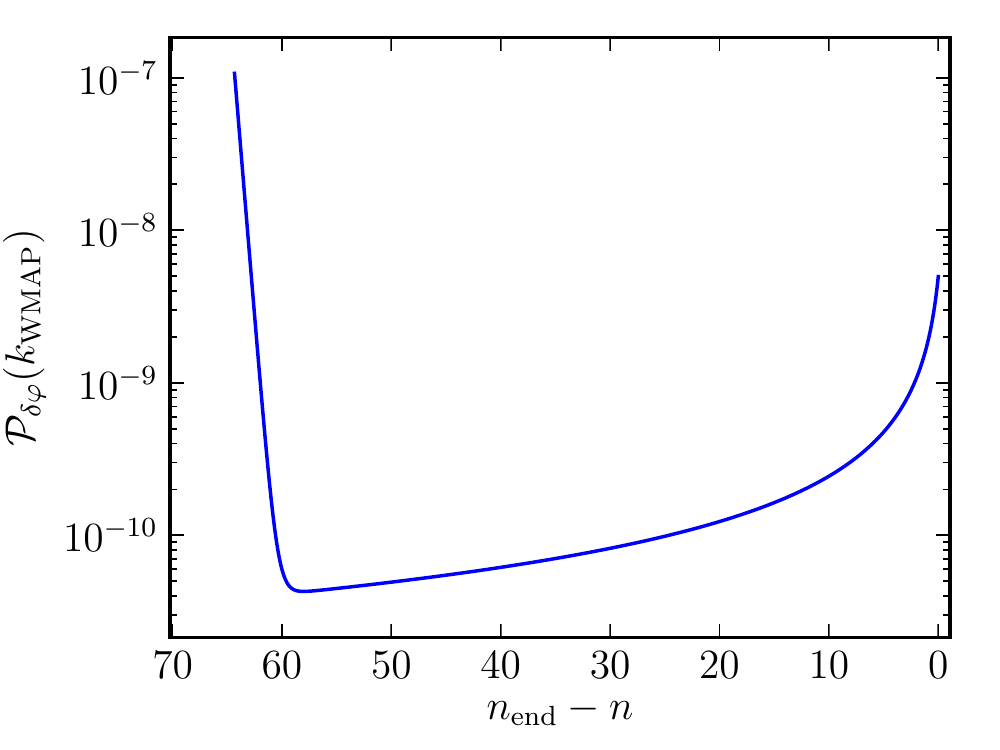}
\label{fig:Pphi-kwmap}
}
\caption{Source term and power spectrum for the WMAP pivot scale $\kwmap$.}
\end{figure}

\begin{figure}
 \subfigure[Comparison of the source term for $\kwmap$ over three
different
ranges with different $\Delta k$s as specified in \eq{eq:Krangedefns}.]{
 \includegraphics[scale=0.8]{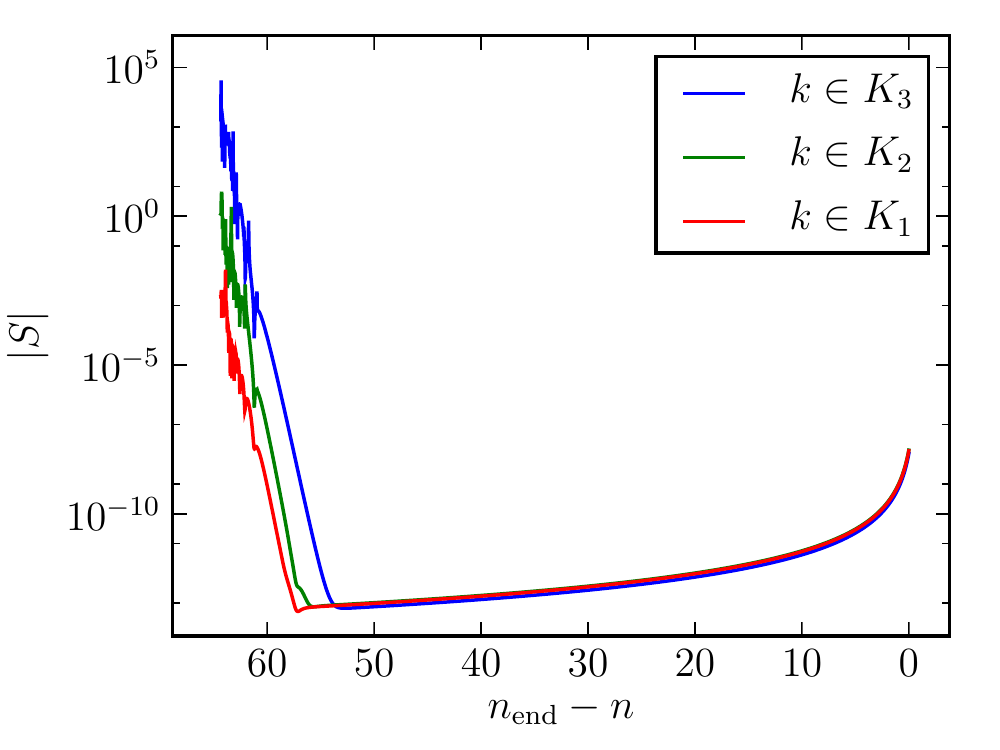}
 \label{fig:src-kwmap-3ranges}
}
\subfigure[The source term for three different $k$ values including the WMAP pivot scale. As $k$
gets larger (scale gets smaller) the source term becomes smaller.]{
 \includegraphics[scale=0.8]{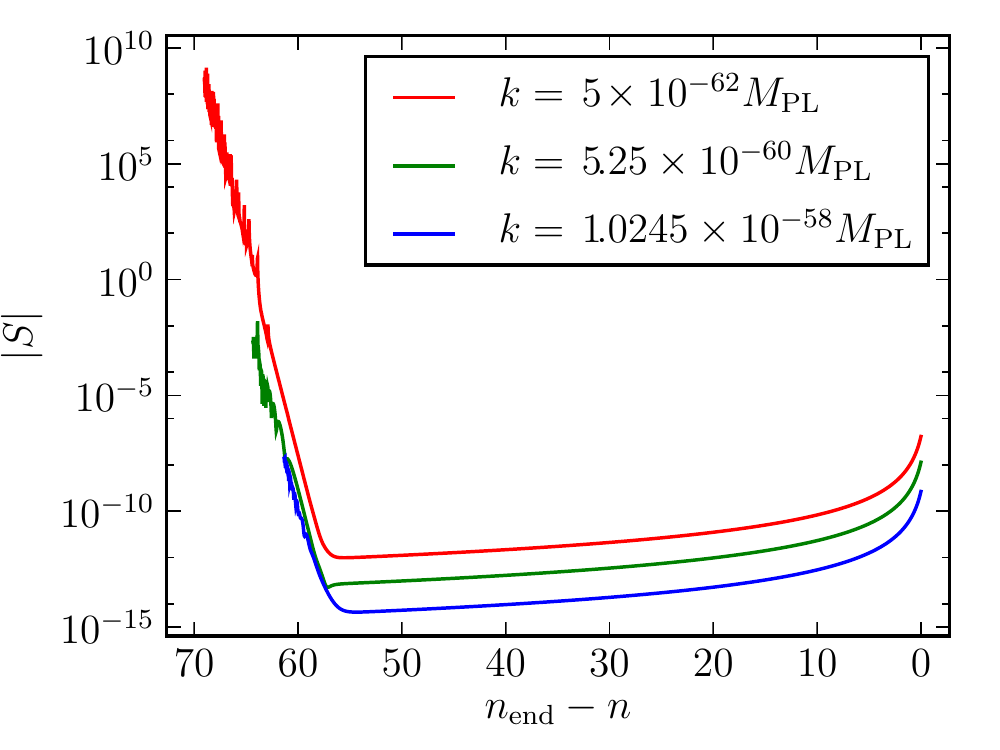}
\label{fig:src-3ks}
}
\caption{Two different comparisons of the source term $S$.}
\label{fig:src-comparisons}
\end{figure}

The source term $S(\kvi)$ is calculated at each time step using the
results of the first order and background runs. This term drives the
production of second order perturbations as shown in
\eqs{KG2_fourier_sr} and
(\ref{KG2_fourier_sr_ntime}). Figure~\ref{fig:src-full} shows the
absolute magnitude of the source term for a single $k$ mode, $\kwmap$,
for all time steps calculated.
Figure~\ref{fig:src-kwmap-3ranges} shows how the source term changes
with the choice of $k$ range.  After horizon crossing the source terms
are approximately equal. Before horizon crossing however there is a
strict hierarchy with the smaller $k$ ranges, $K_1$ and $K_2$, leading to smaller source
contributions.  As stated in Section \ref{sec:codetests}, $\Delta k$
should be at least as large as $\kmin$ in order to reduce the error to
a minimum.

\begin{figure}
\subfigure[The source term (lower blue line) is compared with the $T$ term (upper green
line) as defined in the text for $\kwmap$. The source term is of comparable magnitude at the
beginning of the simulation. ]{
 \includegraphics[scale=0.8]{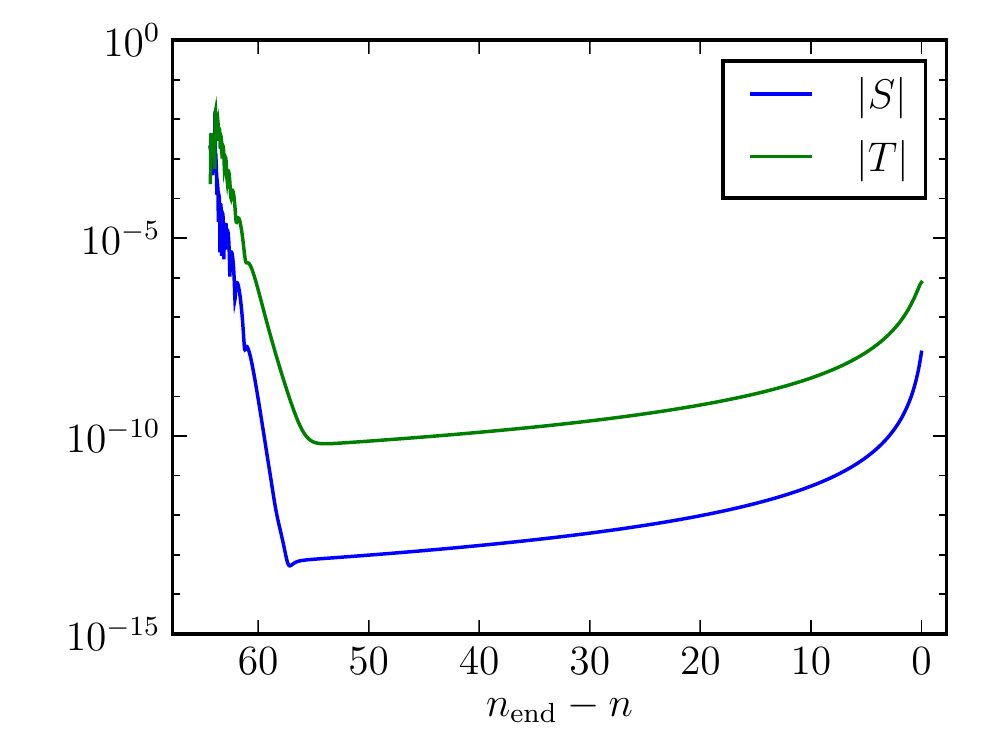}
 \label{fig:src-vs-others}
}
 \subfigure[The quotient of $S$ and $T$ terms for three different $k$ values including
the WMAP pivot scale. Depending on $k$ the source term only dominates at early stages or is
important throughout the evolution.]{
 \includegraphics[scale=0.8]{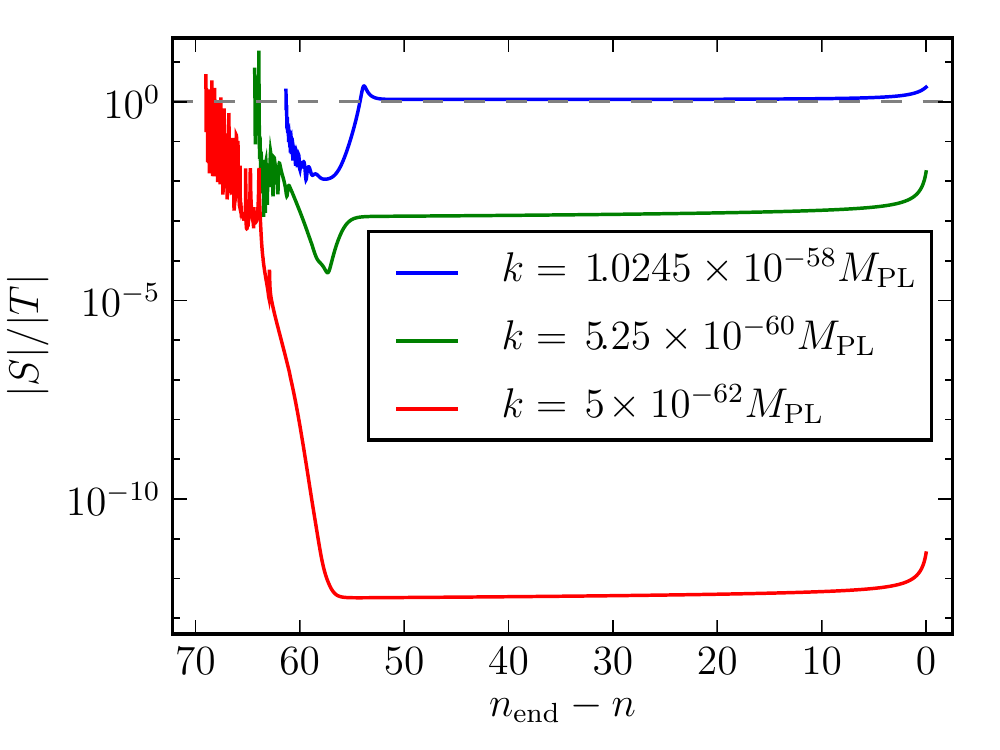}
 \label{fig:s-over-t-3ks}
}
\caption{Source term $S$ compared with $T$.} 
\label{fig:src-3ks-and-over-t}
\end{figure}

The source term is large at early times, and closely follows the form
of the spectrum of the first order perturbations as can be seen from
Figure~\ref{fig:Pphi-kwmap}.
It is useful to compare the magnitude of the source term with the
other terms in the second order evolution equation
(\ref{KG2_fourier_sr_ntime}). If we let $T$ denote the other terms,
\begin{equation}
\label{eq:Tdefn}
 T(\kvi) = \left(3 + \frac{\dot{H}}{H}\right)
\dot{\dvp2}(\kvi)+ \left(\frac{k}{aH}\right)^2\dvp2(\kvi)
+\left(\frac{\Upp}{H^2}-{24 \pi G}(\dot{\vp_{0}})^2\right)
\dvp2(\kvi) \,,
\end{equation}
then Figure~\ref{fig:src-vs-others} shows the absolute magnitude of
both $S$ and $T$.  It is clear that the source term is of comparable
magnitude only early in the simulation.  Figure~\ref{fig:s-over-t-3ks}
shows a comparison of $|S|/|T|$ for three different $k$ values. The larger the $k$ mode the closer
in amplitude $S$ is to the rest of the terms in the ODE.
A priori it is not known
where $S$ will be large for a particular chosen potential and mode but
once determined it could be possible to significantly reduce the time required
for the simulation by only calculating $S$ in the regions where it is
important.
%
%

\begin{figure}
 \subfigure[The $\lambdaphifour$ model (green dashed line) initially has a larger source term but
becomes
smaller than the $\msqphisq$ model as evolution continues. After horizon crossing
the $\lambdaphifour$ term is slightly larger.]{
 \includegraphics[scale=0.8]{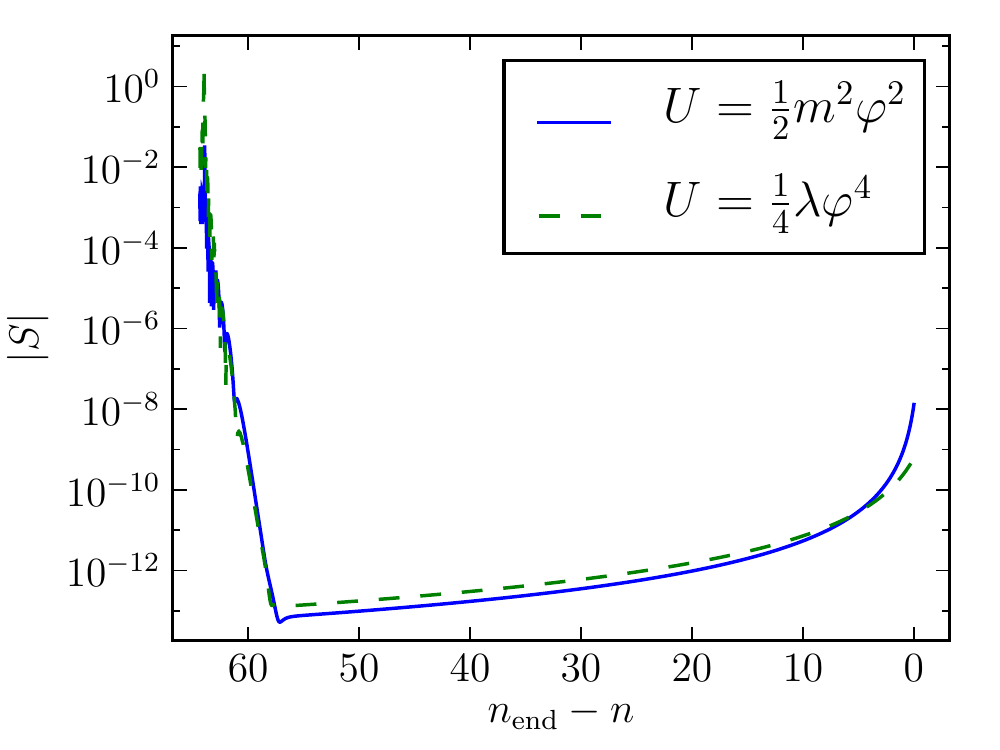}
 \label{fig:src-mvl}
}
\subfigure[The crossover between the models at the early stages of the simulation, before horizon 
crossing.]{
\includegraphics[scale=0.8]{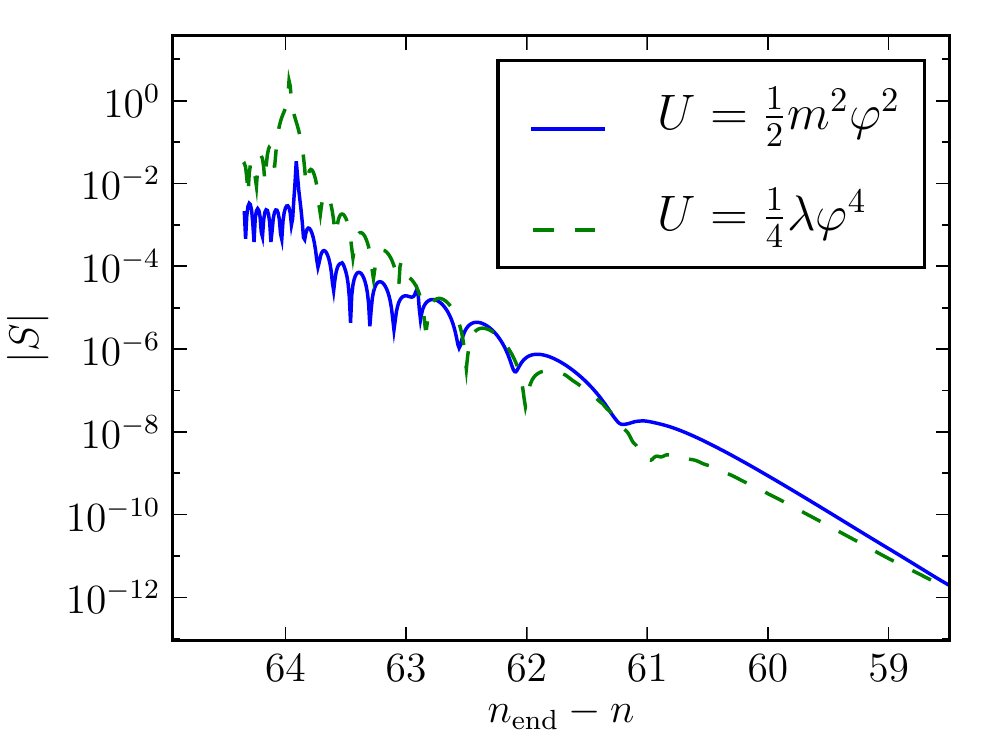}
\label{fig:src-mvl-zoom}
}
\caption{A comparison of the source term for the $\msqphisq$ and
$\lambdaphifour$ models.}
\label{fig:src-mvl-main}
\end{figure}

All the results quoted so far are for
the $\msqphisq$ model. We have also tested the $\lambdaphifour$ model and compared it to
$\msqphisq$. Figure~\ref{fig:src-mvl-main} compares the models for $\kwmap$.
Figure~\ref{fig:src-mvl-zoom} shows how the source term for the $\lambdaphifour$ model is larger
than the one for $\msqphisq$ to begin, but crosses over after a few e-foldings. After horizon
crossing the $\lambdaphifour$ source term is again larger. As the results at first order for both
models are so similar it is to be expected that the second order perturbations would be closely
related. 
%

\begin{figure}
\includegraphics[scale=0.8]{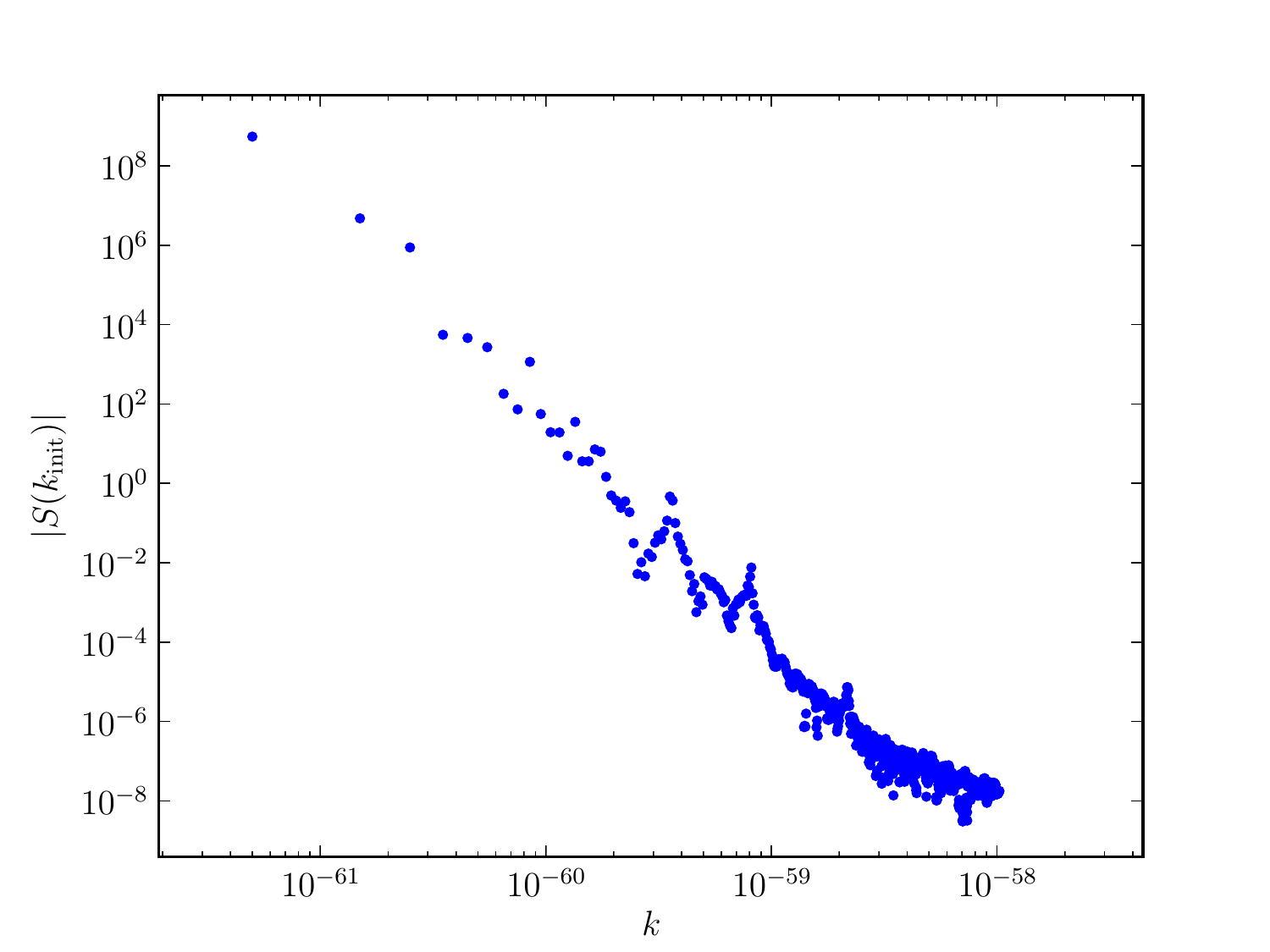}
 \caption{The absolute magnitude of the source term at the initial start time for each $k$ when
$k = aH \times 50$ deep inside the horizon. The results are for the range $K_1=\left[ 1.9\e{-5},
0.039 \right] \Mpc^{-1}= \left[0.5\e{-61}, 1.0245\e{-58}\right]\Mpl$.}
\label{fig:src_kinit}
\end{figure}

In Figure~\ref{fig:src_kinit} the value of $|S|$ at the start of the evolution of $\dvp2$ for
each $k$ mode is shown. The magnitude of the source term is much smaller for larger $k$s (smaller
scales). 
Because the smaller $k$s begin their evolution earlier the relative difference in $|S|$ is not as
pronounced when measured at a single timestep (see for example Figure~\ref{fig:src-3ks}).
It should also be remembered that the magnitude of other terms in the second order ODE is
small for larger $k$s as shown by the ratio $|S|/|T|$ in Figure~\ref{fig:src-3ks-and-over-t}
where $T$ is defined above in \eq{eq:Tdefn}.

\begin{figure}
\includegraphics[scale=0.8]{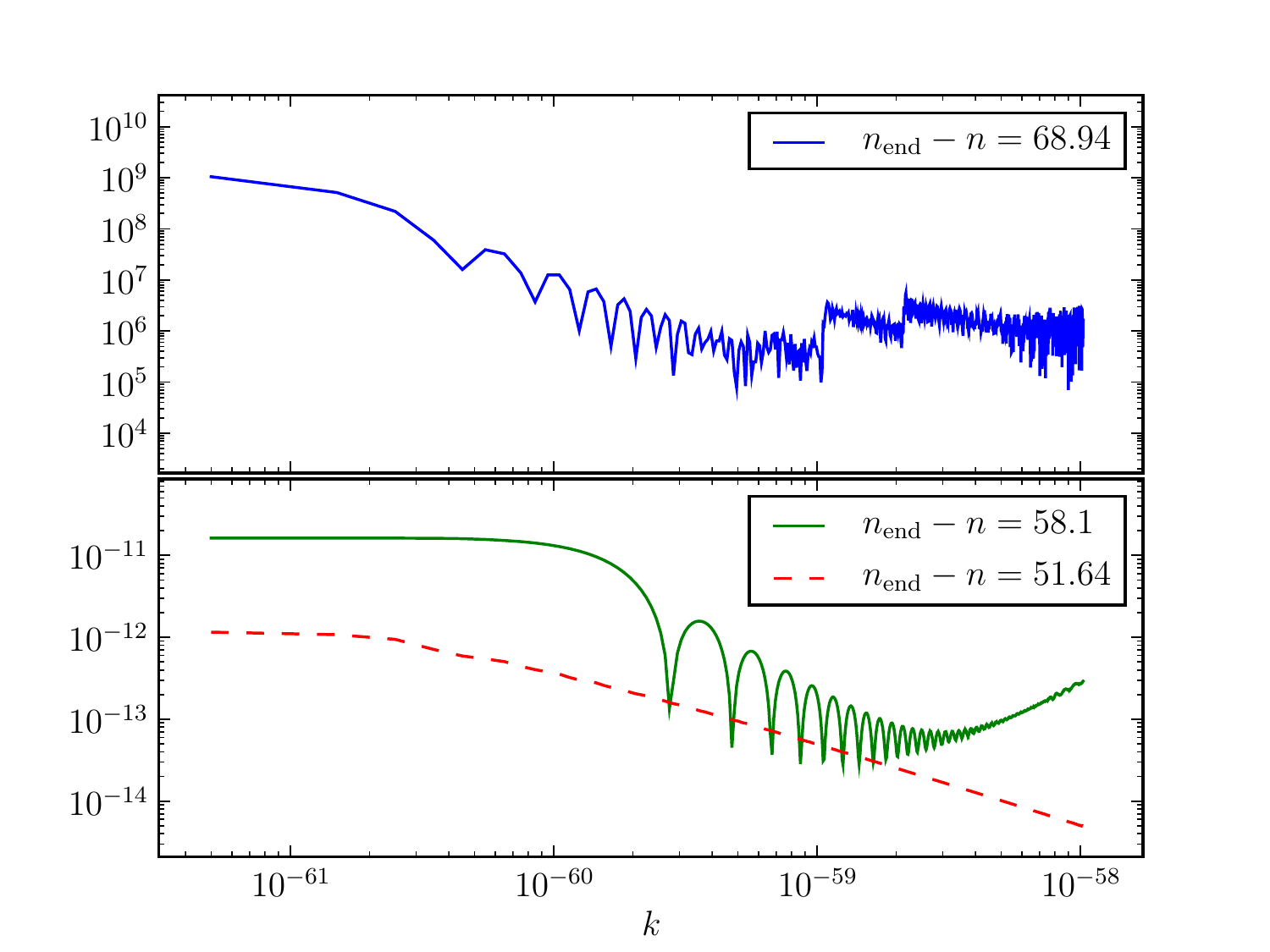}
\caption{The absolute magnitude of the source term for all $ks$ in the range at three different
timesteps: the upper blue line when only the largest modes have been initialised; the middle green
line when all modes have been initialised; and the lower red dashed line when all modes have
exited the horizon. The $k$ range shown here is $K_1=\left[ 1.9\e{-5}, 0.039 \right] \Mpc^{-1}= 
 \left[0.5\e{-61}, 1.0245\e{-58}\right]\Mpl$.}
\label{fig:src_3ns}
\end{figure}

The source term for all $k$s can also be compared for different timesteps. In
Figure~\ref{fig:src_3ns} the upper blue line shows $|S(k)|$ around 69 e-foldings before the end of
inflation when $\dvp2$ has been initialised for only the very smallest $k$ modes. The middle green
line shows $|S|$ when all $\dvp2$ modes have been started. Finally the lower red line plots $|S|$
after all modes have exited the horizon, around 52 e-foldings before the end of inflation.
%

\section{Discussion and conclusion}
\label{sec:disc}

In this paper we have described the numerical solution of the
evolution equations for second order scalar perturbations, using the
closed form of the Klein-Gordon equation, \eq{KG2_fourier_sr}. We
demonstrate that direct calculation of field perturbations beyond
first order using perturbation theory is readily achievable, though
not trivial.

For this first demonstration we have limited ourselves to considering
the slow roll source term in \eq{KG2_fourier_sr} but without imposing
slow roll on the evolution terms of the ODEs. We have investigated two
standard potentials, $\frac{1}{2}m^2\phi^2$ and
$\frac{1}{4}\lambda\phi^4$, to demonstrate the capabilities of the
system. The singularity at $k=0$ which arises as larger and larger
scales are considered is avoided by implementing a cutoff at small
wavenumbers below $\kmin$. This is a pragmatic choice necessary for
the calculation, but as mentioned above there is some evidence that
such a cutoff might also explain lack of power at large scales in the
WMAP data \cite{spergel, Sinha:2005mn, Kim:2009pf}. It is also necessary to
pick a maximum $k$ value, and this choice is dictated by computational
resources and with reference to observationally relevant scales. In
this paper we have used $k$ ranges which are comparable with the
scales observed by WMAP. By comparing the analytical results of the
convolution integral with the numerical calculation, we have chosen
values of the parameters $N_\theta, N_k$ and $\Delta k$ which minimise
the numerical error. The convolution scheme that we have implemented
works best when $\Delta k>\kmin$.

We have shown explicitly that the second order calculations for our chosen
potentials are obtainable once the cut-off for $\kmin$ is
implemented. As expected for these unexceptional potentials in the slowly
rolling regime the magnitude of second order perturbations is extremely
suppressed in comparison with the first order amplitude. We have shown the
evolution of the source term equation during the inflationary regime can be
readily calculated.

There are many possible next steps to improve the program outlined in
Section \ref{sec:numerics}. Chief amongst these is to implement the
full source term equation (\ref{KG2_source_ntime}). Although clearly
more complicated than the slow roll case in \eq{KG2_fourier_sr_ntime}
only two more $\theta$ dependent terms need to be added to $A$--$D$ in
\eq{AtoD}.  For the two test models we have used in this paper, which
are both slowly rolling during inflation, it is not expected that
using the full source equation would result in an appreciably
different outcome until the end of the inflationary phase. Though once
the field has stopped to roll slowly, new observable features might
arise as is indeed the case at first order.

Beyond this the next likely step is to implement a multi-field version of the
system. This would allow the investigation of models that inherently
produce large second order perturbations. In Ref.~\cite{Malik:2006ir} the
Klein-Gordon equation is given for multiple fields and upgrading the
simulation to use these equations is a
straight-forward if lengthy process.

The performance of the numerical simulation could also be improved by
analysing the most time consuming processes and investigating what
optimisations could be implemented. As we have discussed above we have
set $N_k=1025$ for our test runs. This provides good coverage of the
WMAP $k$ range but it is not clear whether it sufficiently
approximates the integral to infinity for the source term.  Currently
we are restricted in our choice of $N_k$ by logistical factors \ie the
running time and memory usage of the code. By optimising the routines
for both memory and speed it is hoped we can extend the maximum value
of $k$ to larger values.

By computing the perturbations to second order we have direct access
to the non-gaussianity of $\dvp{}$.  While useful for the toy models
discussed above (with $\fnl\simeq0$), when used to investigate models
with predictions of large non-linearity parameter $\fnl$ this technique
could yield greater insight into the formation and development of the
non-gaussian contributions by studying the contribution of the different
terms in the source term \eq{KG2_source_ntime}.
It was shown recently that in order to calculate $\fnl$ instead of
using the standard method based on the Lagrangian formalism
\cite{Maldacena:2002vr}, one can instead use the field equations
\cite{Musso:2006pt,Seery:2008qj}. The method presented here will
therefore eventually allow a full numerical calculation of $\fnl$.

In summary, we have demonstrated that numerically solving the closed
Klein-Gordon equation for second order perturbations is possible. We
have used the slow roll version of the source term in this paper, but
hope to extend our work to use the full source soon. The two test
models we have used have been shown to have negligible second order
perturbations in line with analytic results. We have compared the
analytic and numerical solutions for the convolution term and found
them to be in good agreement.

\section*{Acknowledgements}
IH is supported by a STFC and Queen Mary Studentship. The authors
would like to thank Christian Byrnes, Anne Green, Andrew Liddle, Jim Lidsey,
David Seery and Orkan
Umurhan for useful discussions.

\providecommand{\href}[2]{#2}\begingroup\raggedright\endgroup

\end{document}